\documentclass[journal=jacsat,manuscript=article]{achemso}

\usepackage[version=3]{mhchem}
\usepackage{siunitx}
\DeclareSIUnit\angstrom{\protect \text {Å}}

\author{Lucy D. Whalley}
\email{l.whalley@northumbria.ac.uk}
\affiliation[Northumbria University]
{Department of Mathematics, Physics and Electrical Engineering, Northumbria Universiry, Newcastle Upon Tyne, NE1 8ST, UK}

\title{Steric engineering of point defects in lead halide perovskites}

\begin{document}

\begin{tocentry}
\includegraphics[scale=1.0]{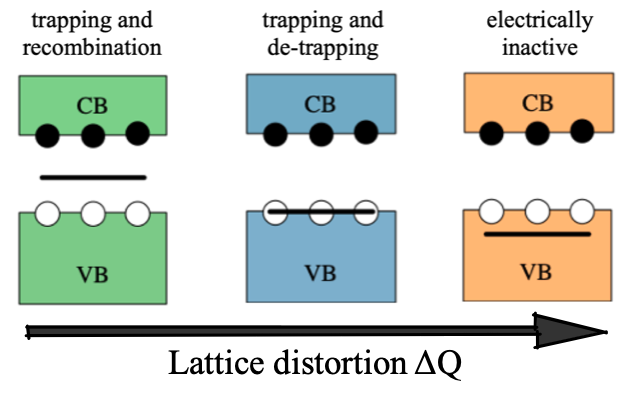}
\end{tocentry}

\abstract{Due to their high photovoltaic efficiency and low-cost synthesis, lead halide perovskites have attracted wide interest for application in new solar cell technologies. The most stable and efficient \ce{ABX_3} perovskite solar cells employ mixed A-site cations, however the impact of cation mixing on carrier trapping and recombination---key processes that limit photovoltaic performance---is not fully understood. Here we analyse non-radiative carrier trapping in the mixed A-cation hybrid halide perovskite \ce{MA_{1-x}Cs_xPbI_3}. By using rigorous first-principles simulations we show that cation mixing leads to a hole trapping rate at the iodine interstitial that is eight orders of magnitude greater than in the single cation system. We demonstrate that the same defect in the same material can display a wide variety of defect activity---from electrically inactive to recombination centre---and, in doing so, resolve conflicting reports in the literature. Finally, we propose a new mechanism in which steric effects can be used to determine the rate of carrier trapping; this is achieved by controlling the phase and dynamical response of the lattice through the A-site composition.
Our findings elucidate crucial links between chemical composition, defect activity and optoelectronic performance, and suggest a general approach that can help to rationalise the development of new crystalline materials with target defect properties.}

\section{Introduction}\label{sec1}

Organic-inorganic lead halide perovskites (OLHPs) have attracted wide interest for their application in optoelectronics. 
Single junction halide perovskite solar cells now exceed 25\% power conversion efficiency \cite{jiang2022surface}, with the most stable and efficient devices employing mixed organic (methylammonium, MA and formamidinium, FA) or inorganic (Cs, Rb) cations on the A-site of the \ce{ABX3} perovskite structure. 
A-site cation engineering is primarily used to improve thermal and chemical stability \cite{Ceratti2021pursuit}; 
the impact of A-site mixing on defect activity is not fully understood,
despite this being crucial for the development of devices with increased efficiency \cite{lee2022rethinking}.

The A-site cation indirectly determines various optoelectronic, transport and defect properties through an influence on the symmetry and dynamic response of the crystal lattice \cite{lee2022rethinking,Egger2018what}. 
Although defect formation and activity is sensitive to the exact system under consideration, the common observation across OLHPs is that halide ions form abundant, mobile point defects which are active in carrier trapping and recombination \cite{du2015density,senocrate2019solid}.
Furthermore, the bonding in OLHPs is relatively weak, leading to an easily distorted (`soft') lattice and large lattice relaxation after carrier capture at a defect site \cite{frost2016what,whalley2021giant}. 

Large lattice relaxation is not commonly observed in all-inorganic optoelectronic materials. It is more akin to what occurs in organic molecular materials \cite{fratini2020charge}, 
where steric engineering through the incorporation of bulky ligands is used to restrict the vibrational modes associated with lattice relaxation after electronic excitation \cite{fratini2020charge,crespo2019exploring}.
Here we propose that a similar approach can be used to rationalise the design of defect-tolerant OLHPs. 
Using first-principles quantum chemical calculations, combined with methods adapted from the field of organic semiconductors, we show that cation mixing in \ce{MA_{1-x}Cs_{x}PbI3} can be used to adjust the hole trapping rate at the iodine interstitial by eight orders of magnitude. Furthermore, using symmetry mode analysis we demonstrate that defect activity can be tuned by controlling phase formation through the steric size of the A-site cations. This provides a new route for engineering defect properties without altering the metal–halide chemistry that is beneficial for photovoltaic performance.

\section{Methods}\label{sec11}

\subsection{Electronic structure calculations}

The underlying electronic structures were calculated using density functional theory (DFT) as implemented in \texttt{VASP} \cite{Kresse1996a} using a plane wave basis set with an energy cutoff of \SI{400}{\electronvolt}. A $2\! \times\!2\!\times\! 2$ gamma centered Monkhorst–Pack mesh was used for the Brillouin zone integration.
The interstitial was placed in a 192-atom supercell. We calculated $\Delta Q$ for the iodine interstitial in a 768 atom unit cell and found it to be converged within \SI{1}{amu\tothe{1/2}\angstrom}. A-site ordering for the mixed cation systems were determined using Special Quasi-random Structures \cite{zunger1990special} as implemented in the \texttt{ICET} code \cite{ICET}.
	
Ground state geometries were found using the PBEsol functional \cite{Perdew2008a} with a force cutoff of \SI{0.01}{\electronvolt\per\angstrom}. The interpolated geometries were generated using custom code (available in an online repository \cite{WhalleyInterpolation}) and the Atomic Simulation Environment \cite{ASE}.
The potential energy surface was calculated using the screened-exchange HSE06 functional \cite{Heyd2003hybrid} with $\alpha = 0.43$ and spin–orbit coupling. The total energy cutoff was \SI{e-5}{\electronvolt}. We used a uniform reduction factor to evaluate HSE06 energies at the gamma point only.  We calculated the neutral defect formation energy in \ce{MAPbI3} using the full $2\! \times\!2\!\times\! 2$ k-point grid and found it to be converged within \SI{0.01}{\electronvolt} per formula unit. We employed a delta self-consistent field approach to constrain the occupation of the defect states near energy level crossings.
The electron–phonon coupling term was derived from wavefunctions calculated at the same level of theory. 

\subsection{Defect properties}

The formation energy of a defect in charge state $q$ is given by
$$E_\mathrm{f}(q) = E_\mathrm{d}(q) - E_\mathrm{b} - \sum_i \mu_i n_i + q(\epsilon_\mathrm{VBM}+E_\mathrm{F}),$$
where $E_\mathrm{d}(q)$ is the total energy of the defect lattice in charge state $q$, $E_\mathrm{b}$ is the total energy of the pristine lattice, $\mu_i$ is the chemical potential of species $i$ and $n_i$ is the number of atoms that are added or removed. $E_\mathrm{d}(q)$, $E_\mathrm{b}$, $\mu_i$ and $\epsilon_\mathrm{VBM}$ were calculated using DFT, as outlined in the previous section. $E_\mathrm{d}(q)$ includes a correction term for charged defects, which was calculated using \texttt{sxdefectalign} with a value of $\bar{\epsilon_0}=22.67$ for the static dielectric constant \cite{sxdefectalign,Brivio2013a}.
More details on the methodology as applied to hybrid halide perovskites have already been published in Reference \cite{whalley2021giant}.  

A quantum mechanical treatment of electron capture was performed using the open-source \texttt{CarrierCapture} package \cite{carriercapture} which builds on the approach outlined in Reference \cite{Alkauskas2014first}.
In this model the carrier capture coefficient for capture from an initial state i to a final state f is given by
$$ C = V\frac{2\pi}{\hbar}gW_\mathrm{if}^2\sum_m\Theta_m\sum_n\vert\langle\chi_\mathrm{im}\vert Q- Q_0\vert \chi_\mathrm{fn}\rangle \vert^2\times \delta(\Delta E+m\hbar\omega_\mathrm{i}- n\hbar\omega_\mathrm{f}), $$
where $\Delta E$ is the total change in energy, $V$ is the supercell volume, $g$ is the energetic degeneracy of the final state, $W_\textrm{if}$ is the electron-phonon coupling matrix element, $\langle\chi_\mathrm{im}\vert Q-Q_0\vert \chi_\mathrm{fn}\rangle$ is the overlap of the vibrational wavefunctions $\chi$, and the Dirac $\delta$ ensures that there is conservation of energy. In practice the Dirac $\delta$ term is replaced by a smearing function; for the calculations in this study this is a gaussian function of width \SI{0.01}{\electronvolt}. $\Theta_m$ is the thermal occupation of the vibrational state $m$. 
The electron-phonon coupling term was calculated using the \texttt{Nonrad} package \cite{nonrad}.
Further details of the methodology can be found in the literature \cite{Alkauskas2014first}.

\subsection{Geometry and symmetry analysis}

Bond lengths and bond angles were analysed using the Atomic Simulation Environment \cite{ASE}. Crystal structures were visualised using \texttt{vesta} \cite{VESTA}. 
\texttt{Isodistort} was used for symmetry mode analysis, with \ce{MAPbI3} in the parent cubic phase $Pm\bar{3}m$ as a reference structure. 
The rotational motion of the MA molecule was not considered in this analysis; all A-sites were modelled as point particles. The phonon mode amplitudes were normalised to the parent cell volume to allow comparison between different compositions. 

\section{Results}\label{sec2}

\subsection{\label{sec:lattice_relaxation}Carrier capture rates with molecular rotations}

\begin{figure*}
\includegraphics[scale=1.0]{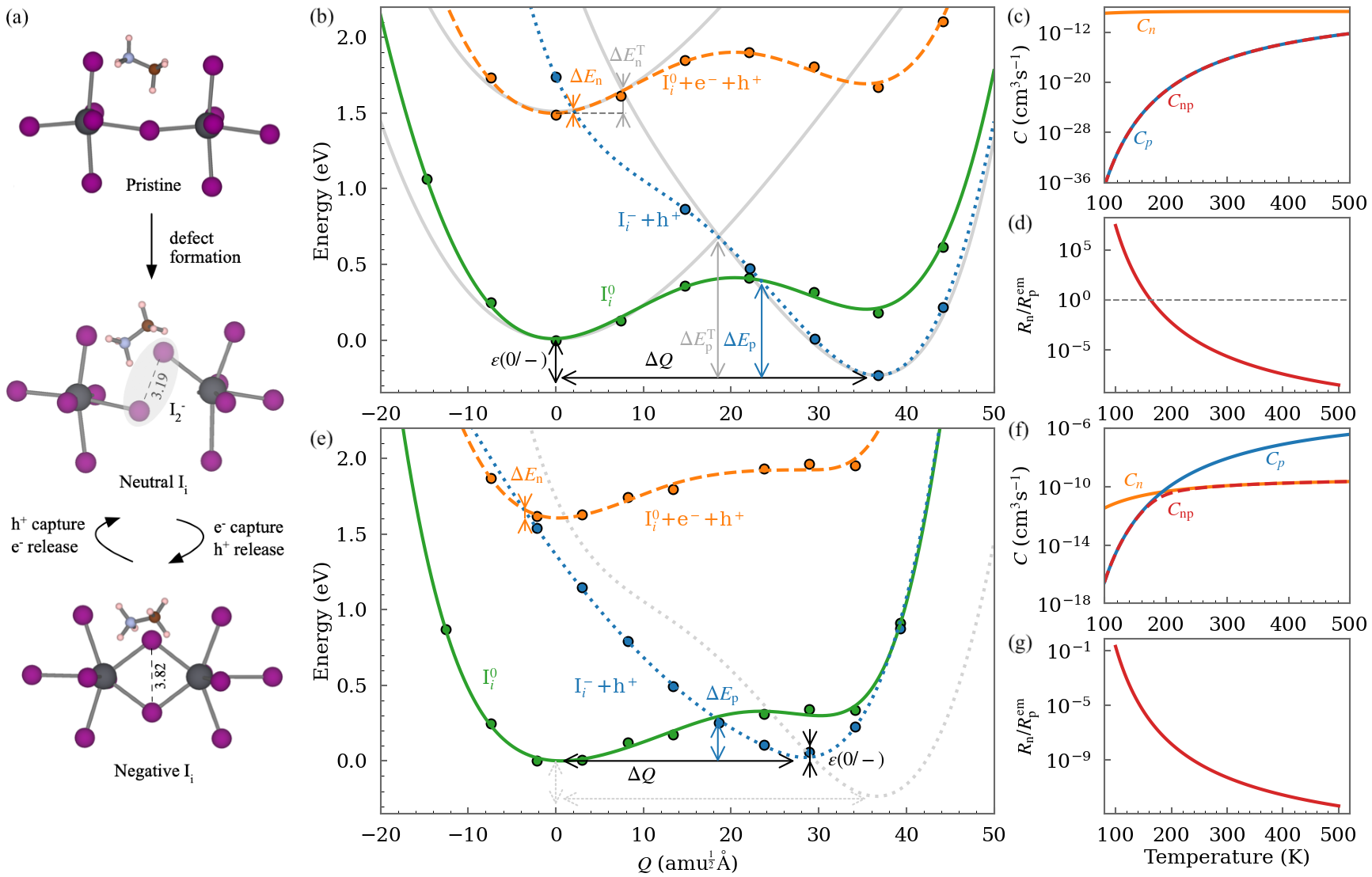}
\caption{\label{fig:CC} Carrier capture processes in single and mixed cation perovskites. Predicted energies are also given in Table S1. (a) Crystal structures for: i) a pristine (defect-free) lead halide perovskite material; ii) the neutral iodine interstitial in a H-centre configuration with a localised hole; iii) the negative iodine interstitial in a split-interstitial configuration. (b) Configuration coordinate diagram for the neutral and negative iodine interstitial in \ce{MAPbI3}. Each scatter point represents a DFT calculated total energy. The solid grey lines are the potential energy surfaces (PES) generated using a lower accuracy interpolation method that describes translations only (no molecular rotations). (c) Nonradiative carrier capture coefficients for the iodine interstitial in \ce{MAPbI3}. Calculated for electron capture at a neutral iodine interstitial (yellow solid line), hole capture at a negative iodine interstitial (blue solid) and electron capture followed by hole capture (red dash) (d) Ratio of electron capture rate (from the conduction band) and hole emission rate (into the valence band) for the neutral iodine interstitial in \ce{MAPbI3}. (e) Configuration coordinate for the neutral and negative iodine interstitial in \ce{MA_{0.875}Cs_{0.125}PbI3}. The grey dot line is the PES of the negative iodine interstitial in \ce{MAPbI3}, given for comparison. (f-g) As in (c-d), but for \ce{MA_{0.875}Cs_{0.125}PbI3}. }
\end{figure*}

We focus our analysis on the iodine interstitial defect in the negative (I$_\mathrm{i}^{-}$) and neutral (I$_\mathrm{i}^{0}$) charge states as these have been found to be most detrimental to solar cell efficiency \cite{Ni2022Evolution}.
The iodine interstitial is a negative-U defect so it follows that I$_\mathrm{i}^{0}$ is metastable \cite{Meggiolaro2018iodine,zhang2020iodine,du2015density,whalley2021giant,whalley2017hcentre}. However I$_\mathrm{i}^{0}$ can still be formed through electron capture at I$_\mathrm{i}^{+}$ or hole capture at I$_\mathrm{i}^{-}$.
As shown in Figure \ref{fig:CC}a, the neutral iodine interstitial I$_\mathrm{i}^0$ bonds with a lattice iodine to produce a molecular I$_{2}^{-}$ H-centre with a trapped hole \cite{whalley2017hcentre}.
After electron capture or hole release the negative charge state I$_\mathrm{i}^{-}$ is formed in a split-interstitial configuration.
This is accompanied by tilting and distortions of the inorganic \ce{PbI6} octahedra to accomodate the I-I bond lengthening. The inorganic structural changes are coupled to MA rotations, similar to the behaviour observed during thermal phase transitions \cite{herz2018how}.

To model the kinetics of charge capture at a defect site we must consider the coupling of electronic and atomic structure. We map the potential energy surface (PES) between two charge states as a function of a collective coordinate 
$$Q=\sqrt{\sum_i m_i \Delta r_{i}^{2}},$$ 
where the sum is over atoms $i$ with mass $m_i$ and a displacement from equilibrium of $\Delta r_i$. For inorganic and hybrid materials the standard procedure is to conduct a linear interpolation between each atomic position of the two equilibrium structures \cite{Alkauskas2014first,kim2019anharmonic,kavanagh2021rapid}, and this is the procedure that has been previously applied to hybrid perovskites \cite{zhang2020iodine,whalley2021giant,Meggiolaro2018iodine,Zhang2022origin}. 
Contributions from the rotation of the molecular cation have previously been ignored as linear interpolation adjusts the intramolecular bond lengths to give unphysical energies \cite{whalley2021giant}. 

We introduce an interpolation method (which we will term `Kabsch interpolation') most commonly used for molecular materials, but transferred here for the first time to a hybrid inorganic-organic material. To interpolate the molecular species we first express atomic positions in the initial and final geometries as a set of vectors, $\mathbf{a}_i$ and $\mathbf{a}_f$ respectively. We use the Kabsch algorithm to calculate an optimal rotation axis $e$ and angle $\theta$ that maps between $\mathbf{a}_i$ and $\mathbf{a}_f$ whilst minimising the root mean squared deviation between each vector pair \cite{Kabsch1976solution}. This allows us to rotate the molecule around $e$ using a linear interpolation of $\theta$. Finally, we combine this with a linear interpolation of both the molecular centre of mass and inorganic framework. This method generates a more accurate energy surface that allows us to consider molecular translations and rotations. Unphysical energies are avoided as the molecule is treated as a rigid object that cannot be deformed.

Figure \ref{fig:CC}b shows configuration coordinate diagrams for the I$^{0}_\mathrm{i}\Leftrightarrow \mathrm{I}^-_\mathrm{i}$ transitions in \ce{MAPbI3}, using standard linear interpolation (in grey) and Kabsch interpolation (in colour). We use the Heyd-Scurseria-Ernzerhof hybrid functional \cite{Heyd2003hybrid} alongside spin-orbit coupling to obtain accurate defect energetics. $Q=0$ corresponds to the equilibrium configuration of I$^{0}_\mathrm{i}$ and $Q=\Delta Q= 
$ \SI{36}{amu\tothe{1/2}\angstrom}  corresponds to that of I$^{-}_\mathrm{i}$. A nonradiative recombination process beings at I$^{0}_\mathrm{i}$ with an electron at the conduction band minimum (CBM) and hole at the valence band maximum (VBM). This is represented with the orange dash line in Figure \ref{fig:CC}b. After electron capture there is a transition to I$^{-}_\mathrm{i}$ (blue dot line); in the semiclassical picture this requires overcoming the energy barrier $ \Delta E_\mathrm{n}$. Subsequent hole capture over the energy barrier $ \Delta E_\mathrm{p}$  to I$^{0}_\mathrm{i}$ (green solid line) completes the recombination cycle. For Kabsch interpolation the energy surface is softened as energy is dissipated through rotations of the MA cation, resulting in a significant reduction of  $ \Delta E_\mathrm{n}$ (\SI{0.15}{\electronvolt} to \SI{0.025}{\electronvolt}) and $\Delta E_\mathrm{p}$ (\SI{0.92}{\electronvolt} to \SI{0.63}{\electronvolt}). 
In semiclassical models the capture rate has an exponential dependence on the ratio of barrier height to $k_\mathrm{B}T$, so changes of $\sim$\SI{100}{\milli\eV}, as seen here, can have a significant impact on defect activity.

For accurate predictions of capture rates, we use a quantum chemical theory to calculate coefficients for electron capture ($C_\mathrm{n}$) and hole capture ($C_\mathrm{p}$) \cite{Alkauskas2014first,kim2019anharmonic}. 
This moves beyond the semiclassical picture to include the strength of coupling between the defect state and CBM ($W^\mathrm{n}_\mathrm{if}$, for electron capture), or the defect state and VBM ($W^\mathrm{p}_\mathrm{if}$, for hole capture). It also considers the vibronic overlap between each PES, thus allowing quantum tunnelling below the classical barrier. 
The capture coefficients determine the capture rate $R$ at a defect. To take electron capture at a neutral defect as an example,
\begin{equation}
    R_n=C_nN_0n,
\end{equation}
where $N_0$ is the neutral defect density and $n$ is the electron density.
We consider electron capture followed by hole capture, the total rate of which is quantified using 
\begin{equation}
    C_\mathrm{np} = \frac{C_\mathrm{n}C_\mathrm{p}}{C_\mathrm{n}+C_\mathrm{p}}.
\end{equation}
We note that this does not correspond to the total rate of non-radiative recombination at the iodine interstitial, as we do not consider non-radiative recombination with hole capture as the initial step.

Figure \ref{fig:CC}c shows that although electron capture is fast at \SI{300}{\kelvin}, the non-radiative recombination process is limited by the slow rate of hole capture. 
Importantly, competing with the first step of this process (electron capture) is hole emission from the localised state associated with I$^0_\mathrm{i}$ (green solid line) to a delocalised state in the valence band (forming I$^{-}_\mathrm{i}$, blue dot line). This process does not require photoexcitation so can happen `in the dark'.
The ratio of the electron capture rate $R_\mathrm{n}$ to the hole emission rate $R^\mathrm{em}_\mathrm{p}$ is given by:
\begin{equation}
\frac{R_\mathrm{n}}{R^\mathrm{em}_\mathrm{p}} = \frac{n C_n}{N_v C_p},
\end{equation}
where $N_\mathrm{v}$ is the density of occupied states in the valence band and $n$ is the electron concentration \cite{das2020what}.
Assuming a hole effective mass value of \SI{0.2}{\emph{m}_e} \cite{Whalley2019Impact} and that the electron concentration is \SI{1e15}{\per\cm\cubed} \cite{Wheeler2017transient}, 
hole emission at the neutral iodine interstitial will occur faster than electron capture at temperatures above \SI{160}{\kelvin} (Figure \ref{fig:CC}d).
It is important to highlight that once the negatively charged iodine interstitial is formed, whether through electron capture or hole emission, it is limited by the slow rate of hole capture ($C_\mathrm{p}=$ \SI{6.0e-17}{\cm\cubed\per\second} at \SI{300}{\kelvin}) and is electrically inactive.

\subsection{\label{sec:lattice_distortions}Carrier capture in mixed A-cation systems}

We now investigate the impact of cation mixing on nonradiative trapping and recombination processes. We consider the mixed cation system \ce{MA_{1-x}Cs_{x}PbI3} and compare this against the control case of \ce{MAPbI3}. 
Initially we focus our analysis on the mixed A-cation system \ce{MA_{0.875}Cs_{0.125}PbI3}, 
which is close to the Cs concentration reported to be optimal for device efficiency \cite{choi2014cesium}.
We find that a H-centre defect is formed with charge localisation around the iodine dimer (Figure S1), indicating that the basic defect activity is comparable to \ce{MAPbI3}.

Figure \ref{fig:CC}e shows that the total lattice relaxation $\Delta Q$ is suppressed through Cs incorporation (\SI{36.8}{amu\tothe{1/2}\angstrom} to \SI{28.9}{amu\tothe{1/2}\angstrom}). 
We note that, despite the reduction, this lattice relaxation is still relatively large as $\Delta Q$ is typically less than \SI{10}{amu\tothe{1/2}\angstrom} for all-inorganic materials \cite{kavanagh2021rapid,kim2019anharmonic}.
Analysis of the defect structure for each charge state shows that the $\Delta Q$ reduction can be primarily attributed to reduced displacements of Pb and I. In particular, the change in Pb-I-Pb bond angle after charge capture is reduced, suggesting that octahedral rotations are suppressed in the mixed cation materials (Table S3).

Figure \ref{fig:CC}e also shows an increase in the neutral to negative charge transition level $\epsilon (0/-)$ (\SI{-0.23}{\electronvolt} to \SI{0.06}{\electronvolt}). 
The charge transition level $\epsilon (q/q^\prime)$ corresponds to the difference in total energy $E_\mathrm{d}$ of defect states $q$ and $q^\prime$ evaluated at their equilibrium configurations $Q$ and $Q^\prime$, respectively, and referenced to the valence band edge $\epsilon_{\mathrm{VBM}}$ of the host material \cite{gallino2010transition}:
\begin{equation}
\epsilon(q/q^\prime) = \frac{E_\mathrm{d}(q^\prime)\rvert_{Q^\prime} - E_\mathrm{d}(q)\rvert_{Q}}{q-q^\prime} -\epsilon_{\mathrm{VBM}}.
\end{equation}
The charge transition level (CTL) is a key parameter for classifying defect activity in the semi-classical defect model. 
For example, defects within a few $K_\mathrm{B}T$ of the band edge typically show single carrier trapping and de-trapping behaviour, whilst `deep' defects towards the middle of a band gap may successively trap both carrier species and form a site for non-radiative recombination \cite{das2020what}.
We note here that the CTL is distinct from the Kohn-Sham defect eigenvalue as the former accounts for structural relaxation after charge
capture or release. The Kohn-Sham defect eigenvalue, often referred to simply as `defect level', has been monitored in previous studies which show that there are large energy fluctuations over \SI{1}{\electronvolt}\cite{Cohen2019breakdown,Wang2022Electron,Chu2020}. 
After cation mixing we find that $\epsilon (0/-)$ CTL increases so that the two charge states become close to thermodynamic equilibrium. As a result, fast hole trapping and de-trapping behaviour is expected.

We find that the shape of the PES is largely unchanged after Cs mixing.
This is especially true at small displacements around the equilibrium structures, where there is an equal softening of the harmonic PES after Cs incorporation for both the negative (\SI{43}{\cm\tothe {-1}}  to \SI{35}{\cm\tothe {-1}}) and neutral (\SI{58}{\cm\tothe {-1}} to \SI{50}{\cm\tothe {-1}}) charge states. For comparison, the \emph{DX}-Centre in GaAs has an effective harmonic frequency of \SI{81}{\cm\tothe {-1}}.
The increase in $\epsilon (0/-)$, combined with a relatively rigid displacement of the $I^{-}_\mathrm{i}$ and $I^{0}_\mathrm{i}$ PES in $E$-$Q$ space, leads to a reduction in the hole capture barrier $\Delta E_\mathrm{p}$ (\SI{0.63}{\electronvolt} to \SI{0.23}{\electronvolt}) and a small increase in the electron barrier height $\Delta E_\mathrm{n}$ (\SI{0.025}{\electronvolt} to \SI{0.045}{\electronvolt}). 
The reduction in $\Delta E_\mathrm{p}$ indicates that there will be an increased rate of hole trapping at the negatively charged iodine interstitial after cation mixing.

To quantify the impact of cation mixing we analyse the same multiphonon capture and recombination processes as outlined in Section \ref{sec:lattice_relaxation}. We find that the defect state couples more strongly with the valence band ($W^\mathrm{p}_\mathrm{if}=$ \SI{0.034}{\eV\,amu \tothe{-1/2}\angstrom \tothe{-1}}) compared to the conduction ($W^\mathrm{n}_\mathrm{if}=$ \SI{0.002}{\eV\,amu \tothe{-1/2}\angstrom \tothe{-1}}); this is expected as the VBM has a larger density of electronic states derived from iodine $p$-orbitals (Figure S2).
This, combined with the reduction in $\Delta E_\mathrm{p}$, results in a carrier recombination process that is more balanced between electron and hole capture, and that shows two clear regimes. At temperatures below \SI{200}{\kelvin} the process is limited by low vibronic overlap between occupied states of the I$^{-}_\mathrm{i}$ and I$^0_\mathrm{i}$ energy surfaces, leading to slow hole capture at I$^{-}_\mathrm{i}$ and a recombination coefficient $C_\mathrm{np}$ that is strongly temperature dependent. At higher temperatures the small $W^\mathrm{n}_\mathrm{if}$ makes electron capture at the I$^{0}_\mathrm{i}$ the limiting process, and $C_\mathrm{np}$ has a reduced dependence on temperature (Figure \ref{fig:CC}f). 

As in the single cation material, we must also consider hole emission from the localised state associated with I$^0_\mathrm{i}$. We find that that the rate of hole emission exceeds the rate of electron capture across the whole temperature range (Figure \ref{fig:CC}g). Furthermore, in contrast to the single cation material, hole trapping is also fast with $C_\mathrm{p}$= \SI{8.4e-9}{\cm\cubed\per\second} at \SI{300}{\kelvin}; an increase by over eight orders of magnitude compared to single cation \ce{MAPbI3}. We conclude that I$_\mathrm{i}$ will form a site for successive hole trapping and de-trapping in this system.

\subsection{\label{sec:lattice_distortions_two}Steric engineering of point defect properties}

\begin{figure}
\includegraphics[scale=1.0]{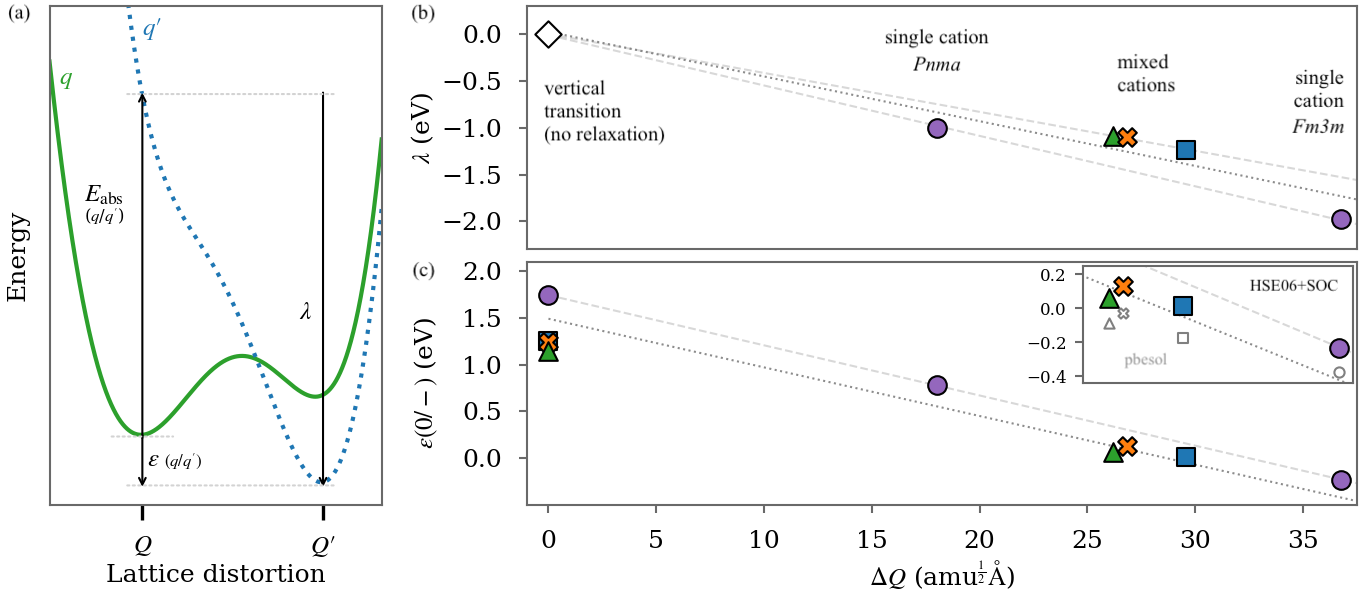}
\caption{\label{fig:EQ_plots} Relationship between defect properties for the \ce{MA_{1-x}Cs_{x}PbI3} series of materials. (a) Schematic showing key defect properties: reorganisation energy $\lambda$, absorption energy $E_\mathrm{abs}(q/q^\prime)$ and charge transition level $\epsilon(q/q^\prime)$ (b) Relationship between reorganisation energy $\lambda$ and $\Delta Q$. The purple circle, green triangle, orange cross and blue square denote $x = 0, 0.125, 0.25$ and 0.5
compositions respectively. Values for the single cation $Pnma$ phase are taken from Reference \cite{zhang2020iodine}. The upper dash line is a first-order polynomial fit to the mixed cation data, the lower dash line is a fit to the single cation data. The dot line is a fit to all scatter points. (c) Relationship between charge transition level $\epsilon (0/-)$ and $\Delta Q$. The dash line is a fit to the single cation data. The dot line is a fit to all scatter points. The inset shows energies calculated with the pbesol exchange-correlation functional (empty scatter points) and the hybrid HSE06 functional with spin-orbit coupling (filled scatter points).}
\end{figure}

The increase in $\epsilon (0/-)$ after cation mixing can be rationalised using the concept of reorganisation energy $\lambda$ from electron transfer theory \cite{Marcus1956theory,bredas2022organic}. 
As shown in Figure \ref{fig:EQ_plots}a, $\lambda$ is the free energy change associated with the relaxation $\Delta Q$ between equilibrium geometries:
\begin{equation}
   \lambda = E_\mathrm{d}(q^\prime)\rvert_{Q^\prime}- E_\mathrm{d}(q^\prime)\rvert_{Q}.
\end{equation}
Note that by definition $\lambda$ is always negative. $\epsilon (q/q^\prime)$ and $\lambda$ are related through the expression $\epsilon (q/q^\prime) = E_\mathrm{abs}(q/q^\prime) \pm \lambda$ \cite{gallino2010transition}, where $E_\mathrm{abs}$ is the Franck-Condon absorption energy for a vertical transition with no change in lattice geometry:
\begin{equation}
    E_\mathrm{abs}(q/q^\prime)  = \frac{E_\mathrm{d}(q^\prime)\rvert_{Q} - E_\mathrm{d}(q)\rvert_{Q}}{q-q^\prime}- \epsilon_\mathrm{VBM}.
\end{equation}
For relaxation after electron capture in \ce{MAPbI3} $\vert\lambda\vert > E_\mathrm{abs}$, so that there is a thermodynamic penalty for re-forming the neutral charge state. 
After Cs incorporation there is a reduction in $\Delta Q$ and, as a result, the magnitude of $\lambda$ is also reduced. This leads to an increase in $\epsilon (0/-)$ so that the two charge states become close to thermodynamic equilibrium.

Reorganisation energy is a key parameter that controls charge mobility in organic electronics and the efficiency of organic photovoltaic devices \cite{shlugerreorganization2018}.
Across a wide range of molecular systems there a correlation between the charge transfer reaction coordinate and $\lambda$; a smaller geometric relaxation is associated with a smaller reorganisation energy. 
This relationship enables charge transport engineering, whereby the size of groups within a molecule and/or molecular packing density is used to determine the extent of lattice relaxation and, following this, $\lambda$ \cite{crespo2019exploring,Ornso2015importance}.

In order to evaluate the extension of steric engineering to lead halide perovskites we begin by investigating the relationship between $\lambda$ and $\Delta Q$ for carrier capture at the iodine interstitial in \ce{MA_{1-x}Cs_{x}PbI3} across a range of stoichiometries ($x=0, 0.125, 0.25$ and 0.5). 
At higher stoichiometries ($x>0.5$) we found that the valence band maximum increases so that there is no localised defect state for hole capture within the band gap, in agreement with previous reports for \ce{CsPbI3} \cite{Jin2020its}.

In Figure \ref{fig:EQ_plots}b we show that after Cs incorporation $\Delta Q$ is reduced relative to the single cation system ($Fm\overline{3}m$ phase) for all compositions. Note that to allow comparison between different A-site compositions we scale $\Delta Q$ by $\sqrt{V_c/V}$ where $V$ is the volume of the defect supercell and $V_c$ is the volume of the control supercell (\ce{MAPbI3} in the $Fm\bar{3}m$ phase). We find that there is a strong linear correlation between $\lambda$ and $\Delta Q$ for the mixed A-cation systems (upper dash line). This analysis includes the point (0,0), as required by the definition of reorganisation energy.

We combine our results with data from the literature at the same level of theory for the single cation system \cite{zhang2020iodine}. 
This enables us to consider three lattice relaxation processes in \ce{MAPbI3}, each corresponding to a different regime of lattice relaxation: i) a vertical transition (no relaxation, $\Delta Q = 0$); ii) relaxation in the $Pnma$ phase (large relaxation, $\Delta Q \sim 20 $); and iii) relaxation in the $Fm\overline{3}m$ phase (giant relaxation, $\Delta Q \sim 35 $). 
Figure \ref{fig:EQ_plots}b confirms a linear correlation between $\lambda$ and $\Delta Q$ for the single cation systems (lower dash line). 

Across all data points (single and mixed cation materials) we find a linear correlation (dot line). 
We also note that for all materials $\vert\lambda\vert$ is in the range \SI{1}{\electronvolt}--\SI{2}{\electronvolt}, which is significantly larger than that typically found for rigid organic molecules (\SI{0.1}{\electronvolt}--\SI{0.2}{\electronvolt}) \cite{shlugerreorganization2018} or point defects in 2D materials ($<$\SI{1}{\electronvolt}) \cite{bertoldo_quantum_2022}.

To understand how $\Delta Q$ may be used to tune the the charge state and activity of I$_\mathrm{i}$ we investigate the relationship between $\epsilon (0/-)$ and $\Delta Q$. 
In Figure \ref{fig:EQ_plots}c we show a negative correlation between $\epsilon (0/-)$ and $\Delta Q$ (dot line), indicating that $\epsilon (0/-)$ can be tuned through variation of $\Delta Q$. As in Figure \ref{fig:EQ_plots}a, if we confine our analysis to a single composition and $E_\mathrm{abs}$ value, the trend becomes stronger (dash line).

The inset in Figure \ref{fig:EQ_plots}c compares $\epsilon (0/-)$ values calculated using the semilocal pbesol exchange-correlation functional without spin-orbit coupling (empty scatter points), and the HSE06 exchange correlation functional with spin-orbit coupling (HSE06-SOC, filled scatter points).
Across all materials we observe a systematic increase in $\epsilon (0/-)$ at the higher HSE06-SOC level of theory. This is due to a shift in the the predicted electronic band edge, as has been observed in previous studies \cite{du2015density}. 
Our results suggest that once the change in $\epsilon (q/q^\prime)$ for a single cation system is known, this can be applied as a correction term to lower accuracy predictions for related mixed cation systems.

Figure \ref{fig:EQ_plots}c provides insight into the significant \SI{0.7}{\electronvolt} discrepancy between previously published values for $\epsilon (0/-)$ \cite{whalley2021giant,zhang2020iodine}.
Our results identify that the source of this discrepancy is the perovskite phase used for modelling, and the influence this has on the predicted $\Delta Q$: whilst
Reference \cite{whalley2021giant} uses the high symmetry $Pm\bar{3}m$ pseudo-cubic phase, Reference \cite{zhang2020iodine} uses the lower symmetry orthorhombic phase, which is formed from condensation of the $M_{3}^{+}$ and $R_4^+$ phonon modes associated with octahedral tilting.
If a subset of the distortions associated with lattice relaxation in the cubic phase is not available in the orthorhombic phase, this will lead to the observed reduction in $\Delta Q$ and increase in $\epsilon (0/-)$.

\begin{figure}
\includegraphics[width=0.5\textwidth]{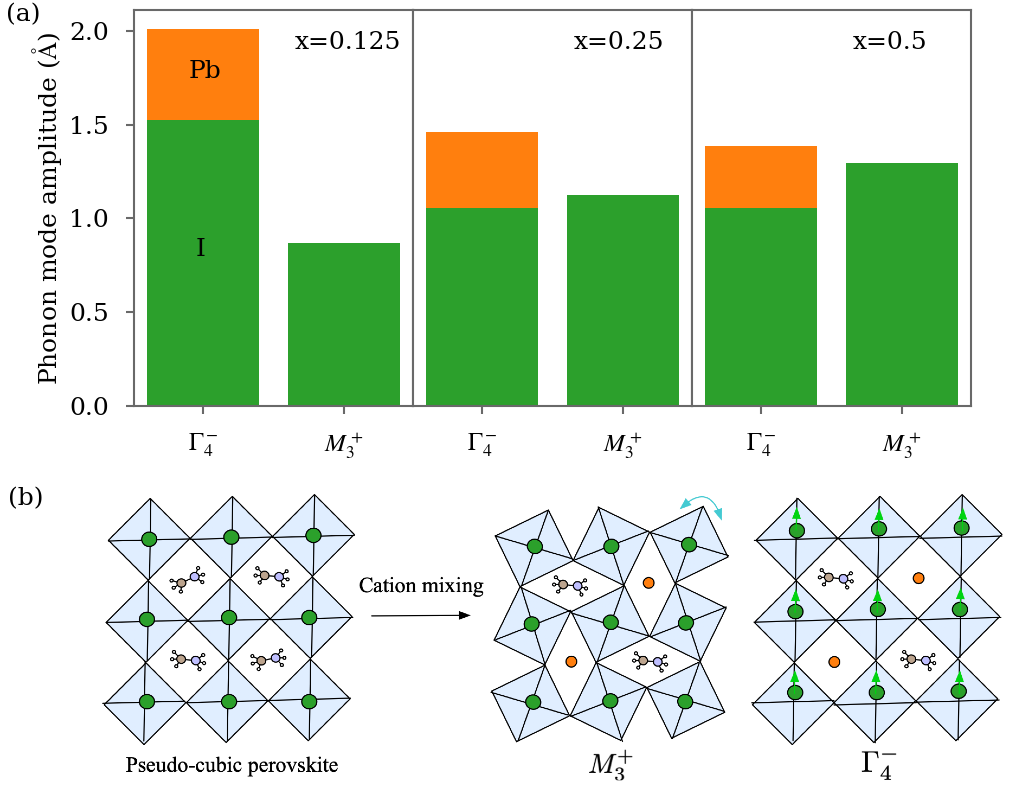}
\caption{\label{fig:SMA} Symmetry mode analysis of \ce{MA_{1-x}Cs_{x}PbI3} with the cubic perovskite phase $Pm\bar{3}m$ used as a reference (parent) structure. a) Amplitude of the two most dominant phonon modes $M_{3}^{+}$ and $\Gamma_{4}^{-}$. Green (Pb) and orange (I) are used to indicate the contribution from each atomic species; the A-site species is not displaced for either mode. b) Schematic of the atomic displacements that correspond to each phonon mode. For easier visualisation of the octahedral tilt patterns the iodine atoms are not shown.}
\end{figure}

Our analysis suggests that a symmetry lowering mechanism might be responsible for the $\Delta Q$ reduction observed after cation mixing.
To test this hypothesis we use symmetry mode analysis \cite{stokes2006isodisplace}. This allows us to decompose the structural changes after A-cation mixing into the normal phonon modes of the single cation cubic structure, giving insight into the specific atomic displacements that are contributing to any lattice distortion. 
Figure \ref{fig:SMA}a shows that for all three mixed cation systems the most dominant phonon modes is $\Gamma_4^-$. This is a polar zone centre displacement corresponding to an off-centering of Pb and I. 
The second most dominant phonon modes is $M_{3}^{+}$. This is a zone boundary mode and a dominant component of thermal phase transitions in perovskite materials \cite{woodward1997octahedral}. It corresponds to rigid rotations of the \ce{PbI6} octahedra, and consequently describes motions of I only. The amplitudes for all phonon modes are given in Figures S3--S5.

Inspection of the $\Gamma_4^-$ polarisation vector shows that this mode is primarily a rigid shift relative to the A-site cation and so does not lead to a reduction in volume.
In contrast, mapping along the $M_{3}^{+}$ mode reduces the metal-halide-metal bonding angle and the cubo-octahedral volume around the A-site. This is in agreement with our measured volumes and Pb-I-Pb bond angle after Cs incorporation (Table S2).
We conclude from this that the reduced steric size of the Cs cation (\SI{1.81}{\angstrom}) compared to MA (\SI{2.70}{\angstrom}) leads to a volume contraction primarily mediated via condensation of the $M_{3}^{+}$ octahedral tilting mode.  

Group theoretical analysis confirms that displacement along $M_{3}^{+}$ reduces the symmetry from cubic to tetragonal ($P4/mbm$), which is in agreement with the structures reported for FA/Cs and FA/MA/Cs mixed cation systems \cite{orri2022high} \cite{prasanna2017band}. 
These experimental observations of mixed A-cation perovskites in the tetragonal phase suggest two important points. Firstly, that this symmetry lowering behaviour is common across other systems where cation mixing leads to a reduction in unit cell volume. Secondly, that any dynamic disorder leading to an effective pseudo-cubic phase is suppressed; the \ce{PbX6} octahedral rotations are `locked-in' and do not time-average to a higher symmetry structure. The latter point is supported by molecular dynamics simulations which show that a low concentration of Cs or Rb in \ce{FAPbI3} suppresses octahedral tilting \cite{Ghosh2018mixed,ghosh2017good}.


\section{Discussion}\label{sec12}

Our calculations show that Cs incorporation leads to a significant increase in the rate of non-radiative trapping and de-trapping at the iodine interstitial. 
At first this may appear to contradict the well-established improvement in performance for mixed cation perovskite materials, which are used in cells with the highest power conversion efficiencies (PCEs) \cite{lee2022rethinking,Hu2018understanding}.
However for high-performance PV materials, both high PCE and good stability (chemical, thermal and mechanical) are a prerequisite.
Figure \ref{fig:CC} suggests that the high PCEs for mixed A-cation perovskites do not derive from a reduced rate of non-radiative trapping or recombination in the bulk (in-grain) material as-synthesised,
but follows from other well-documented factors including increased (photo-)stability \cite{Ceratti2021pursuit,lee2022rethinking,Cho2018high,Niu2018enhancement} and phase purity \cite{saliba2016cesium,Jena2019halide}.

Our results are in agreement the relatively limited experimental characterisation of mixed MA/Cs perovskite materials.
Microwave photoconductivity decay measurements show that Cs-Br incorporation reduces carrier lifetime in MA$_{1- x}$Cs$_x$PbI$_{3-x}$Br$_{x}$ ($x=$ 0.05, 0.1 and 0.15) perovskites films immediately after deposition \cite{Singh2019effect}.
The films were then kept in ambient conditions and the carrier lifetime was measured hourly for nine hours; whilst the carrier lifetime of MAPbI3 films decreased over this time period, the carrier lifetime of the Cs/MA compounds increased.
In a separate study, powder x-ray diffraction shows that incorporation of Cs in MA$_{1-x}$Cs$_{x}$PbBr leads to a reduction in unit cell volume and increased tilting of the \ce{PbBr3} inorganic cage \cite{Premkumar2019impact}. Time-resolved photoluminescence measurements confirm that there is a decrease in the non-radiative recombination lifetime for the MA-rich compounds ($x=$0.2, 0.4) compared to the pure MA compound. 
For increasing Cs content ($x=$0.6, 0.8) the non-radiative lifetime is restored, suggesting that the dominant defect-meditated recombination channel has been removed. 
This supports our calculations showing that there is no active hole trapping state at higher Cs proportion ($x>0.5$).

\begin{figure}
\includegraphics[width=0.5\textwidth]{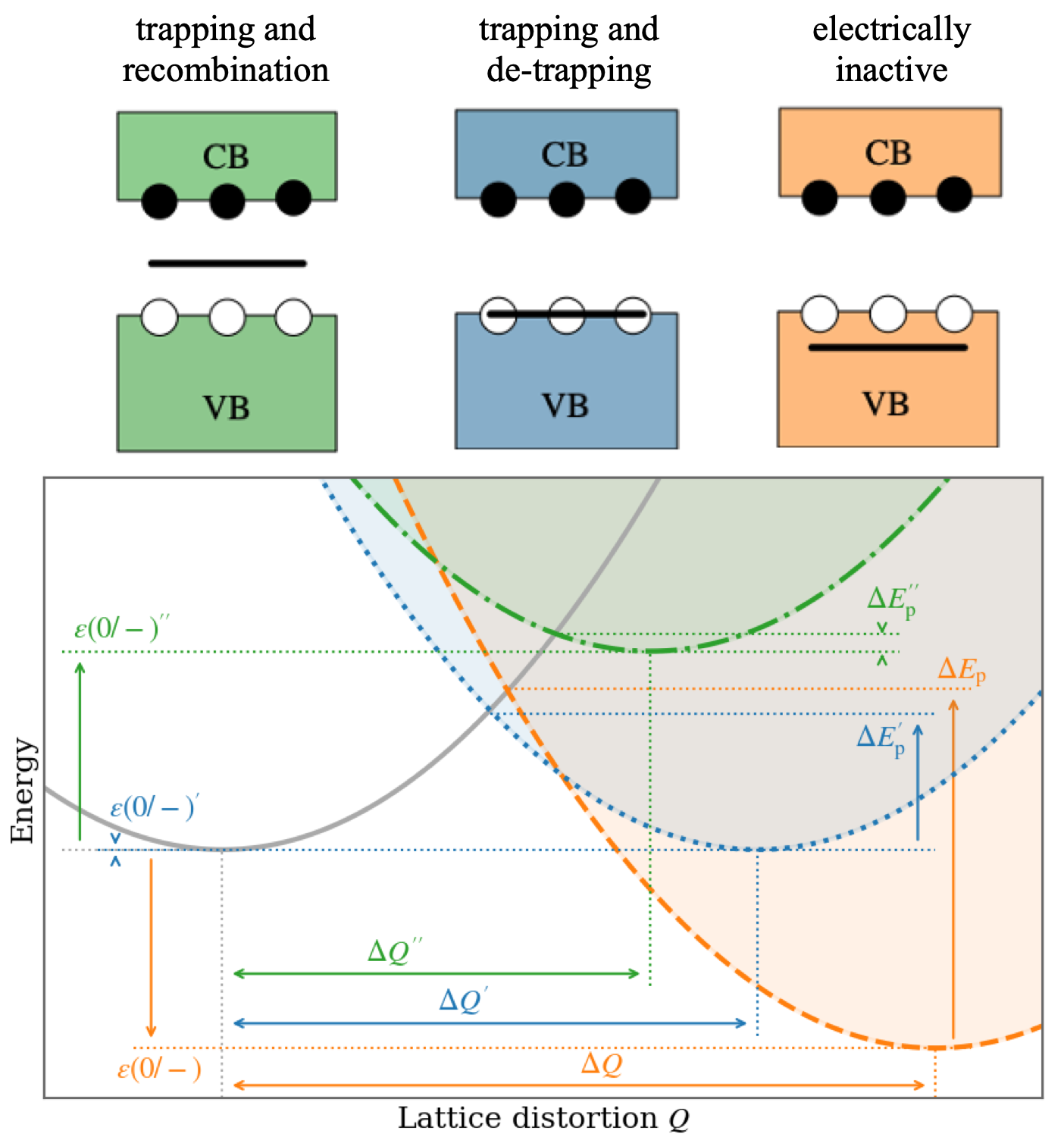}
\caption{\label{fig:schematic} Schematic illustration outlining the impact of steric engineering and crystal phase on I$_\mathrm{i}$ defect activity. $\epsilon (0/-)$ is the neutral to negative charge transition level, $\Delta Q$ is a measure of lattice relaxation between the equilibrium geometries of I$_\mathrm{i}^0$ and I$_\mathrm{i}^-$. The control case of single cation \ce{MAPbI3} in the pseudo-cubic phase shows the largest lattice relaxation after carrier capture, leading to an electrically inactive defect (orange illustrations). A-site mixing with a smaller cation reduces lattice relaxation ($\Delta Q^\prime$) and the hole capture barrier ($\Delta E_\mathrm{p}^\prime$), leading to fast hole trapping and de-trapping (blue). Full symmetry lowering to the orthorhombic phase leads to further reductions in $\Delta Q^{\prime\prime}$ and $\Delta E_\mathrm{p}^{\prime\prime}$, and a charge transition level that is in the electronic band gap (green).}
\end{figure}


Our results demonstrate that CTLs are highly sensitive to $\Delta Q$, and that the same defect in the same material can show a wide variety of defect behaviour. 
As shown schematically in Figure \ref{fig:schematic}, the iodine interstitial can vary from being electrically inactive (with $\epsilon (0/-)$ in the valence band) through being a site for non-radiative hole trapping and de-trapping (with $\epsilon (0/-)< k_\textrm{B}T$), to being a site for non-radiative recombination (with $\epsilon (0/-)$ towards the middle of the band gap).
This conclusion is supported by time resolved photolouminescence measurements of \ce{MAPbI3} which show that regions with greater compressive strain are associated with increased non-radiative decay \cite{Jones2019lattice}.
In addition, first-principles molecular dynamics studies demonstrate that the Kohn-Sham eigenvalues associated with defect states in hybrid halide perovskites can oscillate by as much as \SI{1}{\electronvolt}, reinforcing our finding that defects in this system have an unusually high level of sensitivity to lattice distortions \cite{Cohen2019breakdown,Wang2022Electron,Chu2020}.

Our results also reveal the possibility of tuning defect properties through control of $\Delta Q$, which is a new approach to defect engineering for hybrid and inorganic materials. Heterovalent doping to adjust the Fermi level is firmly established \cite{GrotzCharge2012}, as is adjusting the	chemical potentials of reactants during	growth to increase the formation energy (and thus decrease the concentration) of harmful defects \cite{Walsh2017instilling}.
Instead, we propose isovalent doping (in this case, at the perovskite A-site) to determine the available lattice relaxation pathways ($\Delta Q$), which in turn adjusts the CTL and carrier capture barriers. This approach is enabled by a material system which shows unusually large lattice relaxation and associated re-organisation energies.
In this study we find that the doping-induced volume contraction leads to a symmetry-lowering distortion which `locks in' the accepting phonon modes that are active in non-radiative carrier recombination. However there are other mechanisms beyond this which could used to engineer the available lattice relaxation pathways; for example, applied hydrostatic pressure and epitaxial growth have been shown to induce phase transformations in perovskite materials \cite{Jiao2021strain,Prakash2021self}.

\section{Conclusions}
We have analysed non-radiative carrier trapping in single and mixed A-cation perovskite systems. We have introduced an interpolation method for hybrid materials that describes the coupling between electronic states and molecular rotations, and applied this to \ce{MA_{1-x}Cs_{x}PbI3} for accurate predictions of defect activity. We find that cation mixing leads to a significant increase in the rate of hole trapping and de-trapping compared to the single cation system in the pseudo-cubic phase. 
Importantly, we find a linear relationship between $\epsilon (0/-)$ and $\Delta Q$, demonstrating that the same defect in the same material can display a wide range of defect activity depending on \textit{e.g.} film morphology. 
The reduction in $\Delta Q$ is associated with a symmetry-lowering distortion; in the case of mixed A-site cations this is induced through the decreased steric size of Cs compared to MA.
Furthermore, our results suggest that $\Delta Q$ can be used as an additional parameter to tune carrier trapping activity in crystalline materials. 
This is a general approach for materials that display large lattice relaxation after carrier capture, and may be extended to other hybrid organic-inorganic materials including metal-organic frameworks.

\begin{acknowledgement}
The author thanks Marc Etherington, S\'ean Kavanagh, Jarvist Frost and Aron Walsh for scientific discussions. 
This work used the Oswald High Performance Computing facility operated by Northumbria University (UK). 
Via our membership of the UK's HEC Materials Chemistry Consortium, which is funded by EPSRC (EP/R029431), this work used the ARCHER2 UK National Supercomputing Service (http://archer2.ac.uk).
\end{acknowledgement}

\section{Data Availability Statement}

The custom analysis code and data used to generate the figures and tables in this work are openly available at \url{https://dx.doi.org/10.5281/zenodo.7689415}. Input and output files for the the HSE06-SoC total energy calculations are openly available at \url{https://dx.doi.org/10.17172/NOMAD/2023.02.27-2}. 

\begin{suppinfo}
Supporting Information: DFT calculated energies, charge density distribution, geometric analysis data, electronic density of states, and symmetry mode analysis (PDF)
\end{suppinfo}

\bibliography{bibliography}

\providecommand{\latin}[1]{#1}
\makeatletter
\providecommand{\doi}
  {\begingroup\let\do\@makeother\dospecials
  \catcode`\{=1 \catcode`\}=2 \doi@aux}
\providecommand{\doi@aux}[1]{\endgroup\texttt{#1}}
\makeatother
\providecommand*\mcitethebibliography{\thebibliography}
\csname @ifundefined\endcsname{endmcitethebibliography}
  {\let\endmcitethebibliography\endthebibliography}{}
\begin{mcitethebibliography}{65}
\providecommand*\natexlab[1]{#1}
\providecommand*\mciteSetBstSublistMode[1]{}
\providecommand*\mciteSetBstMaxWidthForm[2]{}
\providecommand*\mciteBstWouldAddEndPuncttrue
  {\def\EndOfBibitem{\unskip.}}
\providecommand*\mciteBstWouldAddEndPunctfalse
  {\let\EndOfBibitem\relax}
\providecommand*\mciteSetBstMidEndSepPunct[3]{}
\providecommand*\mciteSetBstSublistLabelBeginEnd[3]{}
\providecommand*\EndOfBibitem{}
\mciteSetBstSublistMode{f}
\mciteSetBstMaxWidthForm{subitem}{(\alph{mcitesubitemcount})}
\mciteSetBstSublistLabelBeginEnd
  {\mcitemaxwidthsubitemform\space}
  {\relax}
  {\relax}

\bibitem[Jiang \latin{et~al.}(2022)Jiang, Tong, Xian, Kerner, Dunfield, Xiao,
  Scheidt, Kuciauskas, Wang, Hautzinger, Tirawat, Beard, Fenning, Berry,
  Larson, Yan, and Zhu]{jiang2022surface}
Jiang,~Q. \latin{et~al.}  Surface reaction for efficient and stable inverted
  perovskite solar cells. \emph{Nature} \textbf{2022}, \relax
\mciteBstWouldAddEndPunctfalse
\mciteSetBstMidEndSepPunct{\mcitedefaultmidpunct}
{}{\mcitedefaultseppunct}\relax
\EndOfBibitem
\bibitem[Ceratti \latin{et~al.}(2021)Ceratti, Cohen, Tenne, Rakita, Snarski,
  Jasti, Cremonesi, Cohen, Weitman, Rosenhek-Goldian, Kaplan-Ashiri, Bendikov,
  Kalchenko, Elbaum, Potenza, Kronik, Hodes, and Cahen]{Ceratti2021pursuit}
Ceratti,~D.~R. \latin{et~al.}  The pursuit of stability in halide perovskites:
  the monovalent cation and the key for surface and bulk self-healing.
  \emph{Mater. Horiz.} \textbf{2021}, \emph{8}, 1570--1586\relax
\mciteBstWouldAddEndPuncttrue
\mciteSetBstMidEndSepPunct{\mcitedefaultmidpunct}
{\mcitedefaultendpunct}{\mcitedefaultseppunct}\relax
\EndOfBibitem
\bibitem[Lee \latin{et~al.}(2022)Lee, Tan, Seok, Yang, and
  Park]{lee2022rethinking}
Lee,~J.-W.; Tan,~S.; Seok,~S.~I.; Yang,~Y.; Park,~N.-G. Rethinking the {A}
  cation in halide perovskites. \emph{Science} \textbf{2022}, \emph{375},
  eabj1186\relax
\mciteBstWouldAddEndPuncttrue
\mciteSetBstMidEndSepPunct{\mcitedefaultmidpunct}
{\mcitedefaultendpunct}{\mcitedefaultseppunct}\relax
\EndOfBibitem
\bibitem[Egger \latin{et~al.}(2018)Egger, Bera, Cahen, Hodes, Kirchartz,
  Kronik, Lovrincic, Rappe, Reichman, and Yaffe]{Egger2018what}
Egger,~D.~A.; Bera,~A.; Cahen,~D.; Hodes,~G.; Kirchartz,~T.; Kronik,~L.;
  Lovrincic,~R.; Rappe,~A.~M.; Reichman,~D.~R.; Yaffe,~O. What Remains
  Unexplained about the Properties of Halide Perovskites? \emph{Advanced
  Materials} \textbf{2018}, \emph{30}, 1800691\relax
\mciteBstWouldAddEndPuncttrue
\mciteSetBstMidEndSepPunct{\mcitedefaultmidpunct}
{\mcitedefaultendpunct}{\mcitedefaultseppunct}\relax
\EndOfBibitem
\bibitem[Du(2015)]{du2015density}
Du,~M.-H. Density Functional Calculations of Native Defects in \ce{CH3NH3PbI3}:
  Effects of Spin–Orbit Coupling and Self-Interaction Error. \emph{The
  Journal of Physical Chemistry Letters} \textbf{2015}, \emph{6},
  1461--1466\relax
\mciteBstWouldAddEndPuncttrue
\mciteSetBstMidEndSepPunct{\mcitedefaultmidpunct}
{\mcitedefaultendpunct}{\mcitedefaultseppunct}\relax
\EndOfBibitem
\bibitem[Senocrate and Maier(2019)Senocrate, and Maier]{senocrate2019solid}
Senocrate,~A.; Maier,~J. Solid-State Ionics of Hybrid Halide Perovskites.
  \emph{Journal of the American Chemical Society} \textbf{2019}, \emph{141},
  8382--8396\relax
\mciteBstWouldAddEndPuncttrue
\mciteSetBstMidEndSepPunct{\mcitedefaultmidpunct}
{\mcitedefaultendpunct}{\mcitedefaultseppunct}\relax
\EndOfBibitem
\bibitem[Frost and Walsh(2016)Frost, and Walsh]{frost2016what}
Frost,~J.~M.; Walsh,~A. What Is Moving in Hybrid Halide Perovskite Solar Cells?
  \emph{Accounts of Chemical Research} \textbf{2016}, \emph{49}, 528--535\relax
\mciteBstWouldAddEndPuncttrue
\mciteSetBstMidEndSepPunct{\mcitedefaultmidpunct}
{\mcitedefaultendpunct}{\mcitedefaultseppunct}\relax
\EndOfBibitem
\bibitem[Whalley \latin{et~al.}(2021)Whalley, van Gerwen, Frost, Kim, Hood, and
  Walsh]{whalley2021giant}
Whalley,~L.~D.; van Gerwen,~P.; Frost,~J.~M.; Kim,~S.; Hood,~S.~N.; Walsh,~A.
  Giant {H}uang–{R}hys Factor for Electron Capture by the Iodine Intersitial
  in Perovskite Solar Cells. \emph{Journal of the American Chemical Society}
  \textbf{2021}, \emph{143}, 9123--9128\relax
\mciteBstWouldAddEndPuncttrue
\mciteSetBstMidEndSepPunct{\mcitedefaultmidpunct}
{\mcitedefaultendpunct}{\mcitedefaultseppunct}\relax
\EndOfBibitem
\bibitem[Fratini \latin{et~al.}(2020)Fratini, Nikolka, Salleo, Schweicher, and
  Sirringhaus]{fratini2020charge}
Fratini,~S.; Nikolka,~M.; Salleo,~A.; Schweicher,~G.; Sirringhaus,~H. Charge
  transport in high-mobility conjugated polymers and molecular semiconductors.
  \emph{Nature Materials} \textbf{2020}, \emph{19}, 491--502\relax
\mciteBstWouldAddEndPuncttrue
\mciteSetBstMidEndSepPunct{\mcitedefaultmidpunct}
{\mcitedefaultendpunct}{\mcitedefaultseppunct}\relax
\EndOfBibitem
\bibitem[Crespo-Otero \latin{et~al.}(2019)Crespo-Otero, Li, and
  Blancafort]{crespo2019exploring}
Crespo-Otero,~R.; Li,~Q.; Blancafort,~L. Exploring Potential Energy Surfaces
  for Aggregation-Induced Emission—From Solution to Crystal. \emph{Chemistry
  – An Asian Journal} \textbf{2019}, \emph{14}, 700--714\relax
\mciteBstWouldAddEndPuncttrue
\mciteSetBstMidEndSepPunct{\mcitedefaultmidpunct}
{\mcitedefaultendpunct}{\mcitedefaultseppunct}\relax
\EndOfBibitem
\bibitem[Kresse and Furthm{\"{u}}ller(1996)Kresse, and
  Furthm{\"{u}}ller]{Kresse1996a}
Kresse,~G.; Furthm{\"{u}}ller,~J. {Efficient iterative schemes for ab initio
  total-energy calculations using a plane-wave basis set}. \emph{Phys. Rev. B}
  \textbf{1996}, \emph{54}, 11169\relax
\mciteBstWouldAddEndPuncttrue
\mciteSetBstMidEndSepPunct{\mcitedefaultmidpunct}
{\mcitedefaultendpunct}{\mcitedefaultseppunct}\relax
\EndOfBibitem
\bibitem[Zunger \latin{et~al.}(1990)Zunger, Wei, Ferreira, and
  Bernard]{zunger1990special}
Zunger,~A.; Wei,~S.-H.; Ferreira,~L.~G.; Bernard,~J.~E. Special quasirandom
  structures. \emph{Phys. Rev. Lett.} \textbf{1990}, \emph{65}, 353--356\relax
\mciteBstWouldAddEndPuncttrue
\mciteSetBstMidEndSepPunct{\mcitedefaultmidpunct}
{\mcitedefaultendpunct}{\mcitedefaultseppunct}\relax
\EndOfBibitem
\bibitem[Ångqvist \latin{et~al.}(2019)Ångqvist, Muñoz, Rahm, Fransson,
  Durniak, Rozyczko, Rod, and Erhart]{ICET}
Ångqvist,~M.; Muñoz,~W.~A.; Rahm,~J.~M.; Fransson,~E.; Durniak,~C.;
  Rozyczko,~P.; Rod,~T.~H.; Erhart,~P. {ICET} – A Python Library for
  Constructing and Sampling Alloy Cluster Expansions. \emph{Advanced Theory and
  Simulations} \textbf{2019}, \emph{2}, 1900015\relax
\mciteBstWouldAddEndPuncttrue
\mciteSetBstMidEndSepPunct{\mcitedefaultmidpunct}
{\mcitedefaultendpunct}{\mcitedefaultseppunct}\relax
\EndOfBibitem
\bibitem[Perdew \latin{et~al.}(2008)Perdew, Ruzsinszky, Csonka, Vydrov,
  Scuseria, Constantin, Zhou, and Burke]{Perdew2008a}
Perdew,~J.~P.; Ruzsinszky,~A.; Csonka,~G.~I.; Vydrov,~O.~A.; Scuseria,~G.~E.;
  Constantin,~L.~A.; Zhou,~X.; Burke,~K. {Restoring the Density-Gradient
  Expansion for Exchange in Solids and Surfaces}. \emph{Phys. Rev. Lett.}
  \textbf{2008}, \emph{100}, 136406\relax
\mciteBstWouldAddEndPuncttrue
\mciteSetBstMidEndSepPunct{\mcitedefaultmidpunct}
{\mcitedefaultendpunct}{\mcitedefaultseppunct}\relax
\EndOfBibitem
\bibitem[Whalley(2023)]{WhalleyInterpolation}
Whalley,~L.~D. Zenodo repository: {NU-CEM/Kabsch\_interpolation}.
  \textbf{2023}, \relax
\mciteBstWouldAddEndPunctfalse
\mciteSetBstMidEndSepPunct{\mcitedefaultmidpunct}
{}{\mcitedefaultseppunct}\relax
\EndOfBibitem
\bibitem[Larsen \latin{et~al.}(2017)Larsen, Mortensen, Blomqvist, Castelli,
  Christensen, Dułak, Friis, Groves, Hammer, Hargus, Hermes, Jennings, Jensen,
  Kermode, Kitchin, Kolsbjerg, Kubal, Kaasbjerg, Lysgaard, Maronsson, Maxson,
  Olsen, Pastewka, Peterson, Rostgaard, Schiøtz, Schütt, Strange, Thygesen,
  Vegge, Vilhelmsen, Walter, Zeng, and Jacobsen]{ASE}
Larsen,~A.~H. \latin{et~al.}  The atomic simulation environment—a Python
  library for working with atoms. \emph{Journal of Physics: Condensed Matter}
  \textbf{2017}, \emph{29}, 273002\relax
\mciteBstWouldAddEndPuncttrue
\mciteSetBstMidEndSepPunct{\mcitedefaultmidpunct}
{\mcitedefaultendpunct}{\mcitedefaultseppunct}\relax
\EndOfBibitem
\bibitem[Heyd \latin{et~al.}(2003)Heyd, Scuseria, and
  Ernzerhof]{Heyd2003hybrid}
Heyd,~J.; Scuseria,~G.~E.; Ernzerhof,~M. Hybrid functionals based on a screened
  Coulomb potential. \emph{The Journal of Chemical Physics} \textbf{2003},
  \emph{118}, 8207--8215\relax
\mciteBstWouldAddEndPuncttrue
\mciteSetBstMidEndSepPunct{\mcitedefaultmidpunct}
{\mcitedefaultendpunct}{\mcitedefaultseppunct}\relax
\EndOfBibitem
\bibitem[{Sphinx project}()]{sxdefectalign}
{Sphinx project}, Welcome to the SPHInX repository.
  \url{https://sxrepo.mpie.de/}, Accessed: 2019-08-02\relax
\mciteBstWouldAddEndPuncttrue
\mciteSetBstMidEndSepPunct{\mcitedefaultmidpunct}
{\mcitedefaultendpunct}{\mcitedefaultseppunct}\relax
\EndOfBibitem
\bibitem[Brivio \latin{et~al.}(2013)Brivio, Walker, and Walsh]{Brivio2013a}
Brivio,~F.; Walker,~A.~B.; Walsh,~A. {Structural and electronic properties of
  hybrid perovskites for high-efficiency thin-film photovoltaics from
  first-principles}. \emph{APL Mater.} \textbf{2013}, \emph{1}, 042111\relax
\mciteBstWouldAddEndPuncttrue
\mciteSetBstMidEndSepPunct{\mcitedefaultmidpunct}
{\mcitedefaultendpunct}{\mcitedefaultseppunct}\relax
\EndOfBibitem
\bibitem[Kim \latin{et~al.}(2020)Kim, Hood, van Gerwen, Whalley, and
  Walsh]{carriercapture}
Kim,~S.; Hood,~S.~N.; van Gerwen,~P.; Whalley,~L.~D.; Walsh,~A.
  CarrierCapture.jl: Anharmonic Carrier Capture. \emph{J. Open Source Softw.}
  \textbf{2020}, \emph{5}, 2102\relax
\mciteBstWouldAddEndPuncttrue
\mciteSetBstMidEndSepPunct{\mcitedefaultmidpunct}
{\mcitedefaultendpunct}{\mcitedefaultseppunct}\relax
\EndOfBibitem
\bibitem[Alkauskas \latin{et~al.}(2014)Alkauskas, Yan, and Van~de
  Walle]{Alkauskas2014first}
Alkauskas,~A.; Yan,~Q.; Van~de Walle,~C.~G. First-principles theory of
  nonradiative carrier capture via multiphonon emission. \emph{Phys. Rev. B}
  \textbf{2014}, \emph{90}, 075202\relax
\mciteBstWouldAddEndPuncttrue
\mciteSetBstMidEndSepPunct{\mcitedefaultmidpunct}
{\mcitedefaultendpunct}{\mcitedefaultseppunct}\relax
\EndOfBibitem
\bibitem[Turiansky \latin{et~al.}(2021)Turiansky, Alkauskas, Engel, Kresse,
  Wickramaratne, Shen, Dreyer, and Van~de Walle]{nonrad}
Turiansky,~M.~E.; Alkauskas,~A.; Engel,~M.; Kresse,~G.; Wickramaratne,~D.;
  Shen,~J.-X.; Dreyer,~C.~E.; Van~de Walle,~C.~G. Nonrad: {Computing}
  nonradiative capture coefficients from first principles. \emph{Comput. Phys.
  Commun.} \textbf{2021}, \emph{267}, 108056\relax
\mciteBstWouldAddEndPuncttrue
\mciteSetBstMidEndSepPunct{\mcitedefaultmidpunct}
{\mcitedefaultendpunct}{\mcitedefaultseppunct}\relax
\EndOfBibitem
\bibitem[Momma and Izumi(2011)Momma, and Izumi]{VESTA}
Momma,~K.; Izumi,~F. {{\it VESTA3} for three-dimensional visualization of
  crystal, volumetric and morphology data}. \emph{Journal of Applied
  Crystallography} \textbf{2011}, \emph{44}, 1272--1276\relax
\mciteBstWouldAddEndPuncttrue
\mciteSetBstMidEndSepPunct{\mcitedefaultmidpunct}
{\mcitedefaultendpunct}{\mcitedefaultseppunct}\relax
\EndOfBibitem
\bibitem[Ni \latin{et~al.}(2022)Ni, Jiao, Fei, Gu, Xu, Yu, Yang, Deng, Jiang,
  Liu, Yan, and Huang]{Ni2022Evolution}
Ni,~Z.; Jiao,~H.; Fei,~C.; Gu,~H.; Xu,~S.; Yu,~Z.; Yang,~G.; Deng,~Y.;
  Jiang,~Q.; Liu,~Y.; Yan,~Y.; Huang,~J. Evolution of defects during the
  degradation of metal halide perovskite solar cells under reverse bias and
  illumination. \emph{Nature Energy} \textbf{2022}, \emph{7}, 65--73\relax
\mciteBstWouldAddEndPuncttrue
\mciteSetBstMidEndSepPunct{\mcitedefaultmidpunct}
{\mcitedefaultendpunct}{\mcitedefaultseppunct}\relax
\EndOfBibitem
\bibitem[Meggiolaro \latin{et~al.}(2018)Meggiolaro, Motti, Mosconi, Barker,
  Ball, Andrea Riccardo~Perini, Deschler, Petrozza, and
  De~Angelis]{Meggiolaro2018iodine}
Meggiolaro,~D.; Motti,~S.~G.; Mosconi,~E.; Barker,~A.~J.; Ball,~J.; Andrea
  Riccardo~Perini,~C.; Deschler,~F.; Petrozza,~A.; De~Angelis,~F. Iodine
  chemistry determines the defect tolerance of lead-halide perovskites.
  \emph{Energy Environ. Sci.} \textbf{2018}, \emph{11}, 702--713\relax
\mciteBstWouldAddEndPuncttrue
\mciteSetBstMidEndSepPunct{\mcitedefaultmidpunct}
{\mcitedefaultendpunct}{\mcitedefaultseppunct}\relax
\EndOfBibitem
\bibitem[Zhang \latin{et~al.}(2020)Zhang, Turiansky, Shen, and Van~de
  Walle]{zhang2020iodine}
Zhang,~X.; Turiansky,~M.~E.; Shen,~J.-X.; Van~de Walle,~C.~G. Iodine
  interstitials as a cause of nonradiative recombination in hybrid perovskites.
  \emph{Phys. Rev. B} \textbf{2020}, \emph{101}, 140101\relax
\mciteBstWouldAddEndPuncttrue
\mciteSetBstMidEndSepPunct{\mcitedefaultmidpunct}
{\mcitedefaultendpunct}{\mcitedefaultseppunct}\relax
\EndOfBibitem
\bibitem[Whalley \latin{et~al.}(2017)Whalley, Crespo-Otero, and
  Walsh]{whalley2017hcentre}
Whalley,~L.~D.; Crespo-Otero,~R.; Walsh,~A. {H}-Center and {V}-Center Defects
  in Hybrid Halide Perovskites. \emph{ACS Energy Letters} \textbf{2017},
  \emph{2}, 2713--2714\relax
\mciteBstWouldAddEndPuncttrue
\mciteSetBstMidEndSepPunct{\mcitedefaultmidpunct}
{\mcitedefaultendpunct}{\mcitedefaultseppunct}\relax
\EndOfBibitem
\bibitem[Herz(2018)]{herz2018how}
Herz,~L.~M. How Lattice Dynamics Moderate the Electronic Properties of
  Metal-Halide Perovskites. \emph{The Journal of Physical Chemistry Letters}
  \textbf{2018}, \emph{9}, 6853--6863\relax
\mciteBstWouldAddEndPuncttrue
\mciteSetBstMidEndSepPunct{\mcitedefaultmidpunct}
{\mcitedefaultendpunct}{\mcitedefaultseppunct}\relax
\EndOfBibitem
\bibitem[Kim \latin{et~al.}(2019)Kim, Hood, and Walsh]{kim2019anharmonic}
Kim,~S.; Hood,~S.~N.; Walsh,~A. Anharmonic lattice relaxation during
  nonradiative carrier capture. \emph{Phys. Rev. B} \textbf{2019}, \emph{100},
  041202\relax
\mciteBstWouldAddEndPuncttrue
\mciteSetBstMidEndSepPunct{\mcitedefaultmidpunct}
{\mcitedefaultendpunct}{\mcitedefaultseppunct}\relax
\EndOfBibitem
\bibitem[Kavanagh \latin{et~al.}(2021)Kavanagh, Walsh, and
  Scanlon]{kavanagh2021rapid}
Kavanagh,~S.~R.; Walsh,~A.; Scanlon,~D.~O. Rapid Recombination by Cadmium
  Vacancies in \ce{CdTe}. \emph{ACS Energy Letters} \textbf{2021}, \emph{6},
  1392--1398\relax
\mciteBstWouldAddEndPuncttrue
\mciteSetBstMidEndSepPunct{\mcitedefaultmidpunct}
{\mcitedefaultendpunct}{\mcitedefaultseppunct}\relax
\EndOfBibitem
\bibitem[Zhang and Wei(2022)Zhang, and Wei]{Zhang2022origin}
Zhang,~X.; Wei,~S.-H. Origin of Efficiency Enhancement by Lattice Expansion in
  Hybrid-Perovskite Solar Cells. \emph{Phys. Rev. Lett.} \textbf{2022},
  \emph{128}, 136401\relax
\mciteBstWouldAddEndPuncttrue
\mciteSetBstMidEndSepPunct{\mcitedefaultmidpunct}
{\mcitedefaultendpunct}{\mcitedefaultseppunct}\relax
\EndOfBibitem
\bibitem[Kabsch(1976)]{Kabsch1976solution}
Kabsch,~W. {A solution for the best rotation to relate two sets of vectors}.
  \emph{Acta Crystallographica Section A} \textbf{1976}, \emph{32},
  922--923\relax
\mciteBstWouldAddEndPuncttrue
\mciteSetBstMidEndSepPunct{\mcitedefaultmidpunct}
{\mcitedefaultendpunct}{\mcitedefaultseppunct}\relax
\EndOfBibitem
\bibitem[Das \latin{et~al.}(2020)Das, Aguilera, Rau, and
  Kirchartz]{das2020what}
Das,~B.; Aguilera,~I.; Rau,~U.; Kirchartz,~T. What is a deep defect? Combining
  Shockley-Read-Hall statistics with multiphonon recombination theory.
  \emph{Phys. Rev. Materials} \textbf{2020}, \emph{4}, 024602\relax
\mciteBstWouldAddEndPuncttrue
\mciteSetBstMidEndSepPunct{\mcitedefaultmidpunct}
{\mcitedefaultendpunct}{\mcitedefaultseppunct}\relax
\EndOfBibitem
\bibitem[Whalley \latin{et~al.}(2019)Whalley, Frost, Morgan, and
  Walsh]{Whalley2019Impact}
Whalley,~L.~D.; Frost,~J.~M.; Morgan,~B.~J.; Walsh,~A. Impact of nonparabolic
  electronic band structure on the optical and transport properties of
  photovoltaic materials. \emph{Phys. Rev. B} \textbf{2019}, \emph{99},
  085207\relax
\mciteBstWouldAddEndPuncttrue
\mciteSetBstMidEndSepPunct{\mcitedefaultmidpunct}
{\mcitedefaultendpunct}{\mcitedefaultseppunct}\relax
\EndOfBibitem
\bibitem[Wheeler \latin{et~al.}(2017)Wheeler, Bryant, Troughton, Kirchartz,
  Watson, Nelson, and Durrant]{Wheeler2017transient}
Wheeler,~S.; Bryant,~D.; Troughton,~J.; Kirchartz,~T.; Watson,~T.; Nelson,~J.;
  Durrant,~J.~R. Transient Optoelectronic Analysis of the Impact of Material
  Energetics and Recombination Kinetics on the Open-Circuit Voltage of Hybrid
  Perovskite Solar Cells. \emph{The Journal of Physical Chemistry C}
  \textbf{2017}, \emph{121}, 13496--13506\relax
\mciteBstWouldAddEndPuncttrue
\mciteSetBstMidEndSepPunct{\mcitedefaultmidpunct}
{\mcitedefaultendpunct}{\mcitedefaultseppunct}\relax
\EndOfBibitem
\bibitem[Choi \latin{et~al.}(2014)Choi, Jeong, Kim, Kim, Walker, Kim, and
  Kim]{choi2014cesium}
Choi,~H.; Jeong,~J.; Kim,~H.-B.; Kim,~S.; Walker,~B.; Kim,~G.-H.; Kim,~J.~Y.
  Cesium-doped methylammonium lead iodide perovskite light absorber for hybrid
  solar cells. \emph{Nano Energy} \textbf{2014}, \emph{7}, 80--85\relax
\mciteBstWouldAddEndPuncttrue
\mciteSetBstMidEndSepPunct{\mcitedefaultmidpunct}
{\mcitedefaultendpunct}{\mcitedefaultseppunct}\relax
\EndOfBibitem
\bibitem[Gallino \latin{et~al.}(2010)Gallino, Pacchioni, and
  Di~Valentin]{gallino2010transition}
Gallino,~F.; Pacchioni,~G.; Di~Valentin,~C. Transition levels of defect centers
  in ZnO by hybrid functionals and localized basis set approach. \emph{The
  Journal of Chemical Physics} \textbf{2010}, \emph{133}, 144512\relax
\mciteBstWouldAddEndPuncttrue
\mciteSetBstMidEndSepPunct{\mcitedefaultmidpunct}
{\mcitedefaultendpunct}{\mcitedefaultseppunct}\relax
\EndOfBibitem
\bibitem[Cohen \latin{et~al.}(2019)Cohen, Egger, Rappe, and
  Kronik]{Cohen2019breakdown}
Cohen,~A.~V.; Egger,~D.~A.; Rappe,~A.~M.; Kronik,~L. Breakdown of the Static
  Picture of Defect Energetics in Halide Perovskites: The Case of the \ce{Br}
  Vacancy in \ce{CsPbBr3}. \emph{The Journal of Physical Chemistry Letters}
  \textbf{2019}, \emph{10}, 4490--4498\relax
\mciteBstWouldAddEndPuncttrue
\mciteSetBstMidEndSepPunct{\mcitedefaultmidpunct}
{\mcitedefaultendpunct}{\mcitedefaultseppunct}\relax
\EndOfBibitem
\bibitem[Wang \latin{et~al.}(2022)Wang, Chu, Wu, Casanova, Saidi, and
  Prezhdo]{Wang2022Electron}
Wang,~B.; Chu,~W.; Wu,~Y.; Casanova,~D.; Saidi,~W.~A.; Prezhdo,~O.~V.
  Electron-Volt Fluctuation of Defect Levels in Metal Halide Perovskites on a
  100 ps Time Scale. \emph{The Journal of Physical Chemistry Letters}
  \textbf{2022}, \emph{13}, 5946--5952\relax
\mciteBstWouldAddEndPuncttrue
\mciteSetBstMidEndSepPunct{\mcitedefaultmidpunct}
{\mcitedefaultendpunct}{\mcitedefaultseppunct}\relax
\EndOfBibitem
\bibitem[Chu \latin{et~al.}(2020)Chu, Saidi, Zhao, and Prezhdo]{Chu2020}
Chu,~W.; Saidi,~W.~A.; Zhao,~J.; Prezhdo,~O.~V. Soft Lattice and Defect
  Covalency Rationalize Tolerance of $\beta$-CsPbI3 Perovskite Solar Cells to
  Native Defects. \emph{Angewandte Chemie International Edition} \textbf{2020},
  \emph{59}, 6435--6441\relax
\mciteBstWouldAddEndPuncttrue
\mciteSetBstMidEndSepPunct{\mcitedefaultmidpunct}
{\mcitedefaultendpunct}{\mcitedefaultseppunct}\relax
\EndOfBibitem
\bibitem[Marcus(1956)]{Marcus1956theory}
Marcus,~R.~A. On the Theory of Oxidation‐Reduction Reactions Involving
  Electron Transfer. \emph{The Journal of Chemical Physics} \textbf{1956},
  \emph{24}, 966--978\relax
\mciteBstWouldAddEndPuncttrue
\mciteSetBstMidEndSepPunct{\mcitedefaultmidpunct}
{\mcitedefaultendpunct}{\mcitedefaultseppunct}\relax
\EndOfBibitem
\bibitem[Brédas \latin{et~al.}(2002)Brédas, Calbert, da~Silva~Filho, and
  Cornil]{bredas2022organic}
Brédas,~J.~L.; Calbert,~J.~P.; da~Silva~Filho,~D.~A.; Cornil,~J. Organic
  semiconductors: A theoretical characterization of the basic parameters
  governing charge transport. \emph{Proceedings of the National Academy of
  Sciences} \textbf{2002}, \emph{99}, 5804--5809\relax
\mciteBstWouldAddEndPuncttrue
\mciteSetBstMidEndSepPunct{\mcitedefaultmidpunct}
{\mcitedefaultendpunct}{\mcitedefaultseppunct}\relax
\EndOfBibitem
\bibitem[Shluger and Grutter(2018)Shluger, and
  Grutter]{shlugerreorganization2018}
Shluger,~A.~L.; Grutter,~P. Reorganization takes energy. \emph{Nature
  Nanotechnology} \textbf{2018}, \emph{13}, 360--361\relax
\mciteBstWouldAddEndPuncttrue
\mciteSetBstMidEndSepPunct{\mcitedefaultmidpunct}
{\mcitedefaultendpunct}{\mcitedefaultseppunct}\relax
\EndOfBibitem
\bibitem[{\O}rns{\o} \latin{et~al.}(2015){\O}rns{\o}, J\'{o}nsson, Jacobsen,
  and Thygesen]{Ornso2015importance}
{\O}rns{\o},~K.~B.; J\'{o}nsson,~E.~O.; Jacobsen,~K.~W.; Thygesen,~K.~S.
  Importance of the Reorganization Energy Barrier in Computational Design of
  Porphyrin-Based Solar Cells with Cobalt-Based Redox Mediators. \emph{The
  Journal of Physical Chemistry C} \textbf{2015}, \emph{119},
  12792--12800\relax
\mciteBstWouldAddEndPuncttrue
\mciteSetBstMidEndSepPunct{\mcitedefaultmidpunct}
{\mcitedefaultendpunct}{\mcitedefaultseppunct}\relax
\EndOfBibitem
\bibitem[Jin \latin{et~al.}(2020)Jin, Debroye, Keshavarz, Scheblykin,
  Roeffaers, Hofkens, and Steele]{Jin2020its}
Jin,~H.; Debroye,~E.; Keshavarz,~M.; Scheblykin,~I.~G.; Roeffaers,~M. B.~J.;
  Hofkens,~J.; Steele,~J.~A. It{'}s a trap! On the nature of localised states
  and charge trapping in lead halide perovskites. \emph{Mater. Horiz.}
  \textbf{2020}, \emph{7}, 397--410\relax
\mciteBstWouldAddEndPuncttrue
\mciteSetBstMidEndSepPunct{\mcitedefaultmidpunct}
{\mcitedefaultendpunct}{\mcitedefaultseppunct}\relax
\EndOfBibitem
\bibitem[Bertoldo \latin{et~al.}(2022)Bertoldo, Ali, Manti, and
  Thygesen]{bertoldo_quantum_2022}
Bertoldo,~F.; Ali,~S.; Manti,~S.; Thygesen,~K.~S. Quantum point defects in {2D}
  materials - the {QPOD} database. \emph{npj Computational Materials}
  \textbf{2022}, \emph{8}, 56\relax
\mciteBstWouldAddEndPuncttrue
\mciteSetBstMidEndSepPunct{\mcitedefaultmidpunct}
{\mcitedefaultendpunct}{\mcitedefaultseppunct}\relax
\EndOfBibitem
\bibitem[Stokes \latin{et~al.}(2006)Stokes, Hatch, Campbell, and
  Tanner]{stokes2006isodisplace}
Stokes,~H.~T.; Hatch,~D.~M.; Campbell,~B.~J.; Tanner,~D.~E. {ISODISPLACE}: a
  web-based tool for exploring structural distortions. \emph{Journal of Applied
  Crystallography} \textbf{2006}, \emph{39}, 607--614\relax
\mciteBstWouldAddEndPuncttrue
\mciteSetBstMidEndSepPunct{\mcitedefaultmidpunct}
{\mcitedefaultendpunct}{\mcitedefaultseppunct}\relax
\EndOfBibitem
\bibitem[Woodward(1997)]{woodward1997octahedral}
Woodward,~P.~M. Octahedral Tilting in Perovskites. II. Structure Stabilizing
  Forces. \emph{Acta Crystallographica Section B} \textbf{1997}, \emph{53},
  44--66\relax
\mciteBstWouldAddEndPuncttrue
\mciteSetBstMidEndSepPunct{\mcitedefaultmidpunct}
{\mcitedefaultendpunct}{\mcitedefaultseppunct}\relax
\EndOfBibitem
\bibitem[Orri \latin{et~al.}()Orri, Doherty, Johnstone, Collins, Simons,
  Midgley, Ducati, and Stranks]{orri2022high}
Orri,~J.~F.; Doherty,~T.~A.; Johnstone,~D.; Collins,~S.~M.; Simons,~H.;
  Midgley,~P.~A.; Ducati,~C.; Stranks,~S.~D. Unveiling the interaction
  mechanisms of electron and X-ray radiation with halide perovskite
  semiconductors using scanning nano-probe diffraction. \emph{Advanced
  Materials} \emph{34}, 2200383\relax
\mciteBstWouldAddEndPuncttrue
\mciteSetBstMidEndSepPunct{\mcitedefaultmidpunct}
{\mcitedefaultendpunct}{\mcitedefaultseppunct}\relax
\EndOfBibitem
\bibitem[Prasanna \latin{et~al.}(2017)Prasanna, Gold-Parker, Leijtens, Conings,
  Babayigit, Boyen, Toney, and McGehee]{prasanna2017band}
Prasanna,~R.; Gold-Parker,~A.; Leijtens,~T.; Conings,~B.; Babayigit,~A.;
  Boyen,~H.-G.; Toney,~M.~F.; McGehee,~M.~D. Band Gap Tuning via Lattice
  Contraction and Octahedral Tilting in Perovskite Materials for Photovoltaics.
  \emph{Journal of the American Chemical Society} \textbf{2017}, \emph{139},
  11117--11124\relax
\mciteBstWouldAddEndPuncttrue
\mciteSetBstMidEndSepPunct{\mcitedefaultmidpunct}
{\mcitedefaultendpunct}{\mcitedefaultseppunct}\relax
\EndOfBibitem
\bibitem[Ghosh \latin{et~al.}(2018)Ghosh, Smith, Walker, and
  Islam]{Ghosh2018mixed}
Ghosh,~D.; Smith,~A.~R.; Walker,~A.~B.; Islam,~M.~S. Mixed {A}-Cation
  Perovskites for Solar Cells: Atomic-Scale Insights Into Structural
  Distortion, Hydrogen Bonding, and Electronic Properties. \emph{Chemistry of
  Materials} \textbf{2018}, \emph{30}, 5194--5204\relax
\mciteBstWouldAddEndPuncttrue
\mciteSetBstMidEndSepPunct{\mcitedefaultmidpunct}
{\mcitedefaultendpunct}{\mcitedefaultseppunct}\relax
\EndOfBibitem
\bibitem[Ghosh \latin{et~al.}(2017)Ghosh, Walsh~Atkins, Islam, Walker, and
  Eames]{ghosh2017good}
Ghosh,~D.; Walsh~Atkins,~P.; Islam,~M.~S.; Walker,~A.~B.; Eames,~C. Good
  Vibrations: Locking of Octahedral Tilting in Mixed-Cation Iodide Perovskites
  for Solar Cells. \emph{ACS Energy Letters} \textbf{2017}, \emph{2},
  2424--2429\relax
\mciteBstWouldAddEndPuncttrue
\mciteSetBstMidEndSepPunct{\mcitedefaultmidpunct}
{\mcitedefaultendpunct}{\mcitedefaultseppunct}\relax
\EndOfBibitem
\bibitem[Hu \latin{et~al.}(2018)Hu, Hutter, Rieder, Grill, Hanisch, Aygüler,
  Hufnagel, Handloser, Bein, Hartschuh, Tvingstedt, Dyakonov, Baumann,
  Savenije, Petrus, and Docampo]{Hu2018understanding}
Hu,~Y. \latin{et~al.}  Understanding the Role of Cesium and Rubidium Additives
  in Perovskite Solar Cells: Trap States, Charge Transport, and Recombination.
  \emph{Advanced Energy Materials} \textbf{2018}, \emph{8}, 1703057\relax
\mciteBstWouldAddEndPuncttrue
\mciteSetBstMidEndSepPunct{\mcitedefaultmidpunct}
{\mcitedefaultendpunct}{\mcitedefaultseppunct}\relax
\EndOfBibitem
\bibitem[Cho \latin{et~al.}(2018)Cho, Kim, Wolf, Kim, Yun, Jeong, Sadhanala,
  Venugopalan, Choi, Lee, Friend, and Lee]{Cho2018high}
Cho,~H.; Kim,~J.~S.; Wolf,~C.; Kim,~Y.-H.; Yun,~H.~J.; Jeong,~S.-H.;
  Sadhanala,~A.; Venugopalan,~V.; Choi,~J.~W.; Lee,~C.-L.; Friend,~R.~H.;
  Lee,~T.-W. High-Efficiency Polycrystalline Perovskite Light-Emitting Diodes
  Based on Mixed Cations. \emph{ACS Nano} \textbf{2018}, \emph{12}, 2883--2892,
  PMID: 29494128\relax
\mciteBstWouldAddEndPuncttrue
\mciteSetBstMidEndSepPunct{\mcitedefaultmidpunct}
{\mcitedefaultendpunct}{\mcitedefaultseppunct}\relax
\EndOfBibitem
\bibitem[Niu \latin{et~al.}(2017)Niu, Li, Li, Liang, and
  Wang]{Niu2018enhancement}
Niu,~G.; Li,~W.; Li,~J.; Liang,~X.; Wang,~L. Enhancement of thermal stability
  for perovskite solar cells through cesium doping. \emph{RSC Adv.}
  \textbf{2017}, \emph{7}, 17473--17479\relax
\mciteBstWouldAddEndPuncttrue
\mciteSetBstMidEndSepPunct{\mcitedefaultmidpunct}
{\mcitedefaultendpunct}{\mcitedefaultseppunct}\relax
\EndOfBibitem
\bibitem[Saliba \latin{et~al.}(2016)Saliba, Matsui, Seo, Domanski,
  Correa-Baena, Nazeeruddin, Zakeeruddin, Tress, Abate, Hagfeldt, and
  Grätzel]{saliba2016cesium}
Saliba,~M.; Matsui,~T.; Seo,~J.-Y.; Domanski,~K.; Correa-Baena,~J.-P.;
  Nazeeruddin,~M.~K.; Zakeeruddin,~S.~M.; Tress,~W.; Abate,~A.; Hagfeldt,~A.;
  Grätzel,~M. Cesium-containing triple cation perovskite solar cells: improved
  stability{,} reproducibility and high efficiency. \emph{Energy Environ. Sci.}
  \textbf{2016}, \emph{9}, 1989--1997\relax
\mciteBstWouldAddEndPuncttrue
\mciteSetBstMidEndSepPunct{\mcitedefaultmidpunct}
{\mcitedefaultendpunct}{\mcitedefaultseppunct}\relax
\EndOfBibitem
\bibitem[Jena \latin{et~al.}(2019)Jena, Kulkarni, and Miyasaka]{Jena2019halide}
Jena,~A.~K.; Kulkarni,~A.; Miyasaka,~T. Halide Perovskite Photovoltaics:
  Background, Status, and Future Prospects. \emph{Chemical Reviews}
  \textbf{2019}, \emph{119}, 3036--3103\relax
\mciteBstWouldAddEndPuncttrue
\mciteSetBstMidEndSepPunct{\mcitedefaultmidpunct}
{\mcitedefaultendpunct}{\mcitedefaultseppunct}\relax
\EndOfBibitem
\bibitem[Singh \latin{et~al.}(2019)Singh, Chouhan, and
  Avasthi]{Singh2019effect}
Singh,~A.; Chouhan,~A.~S.; Avasthi,~S. Effect of methylamine vapor exposure and
  ambient ageing on Cs$_{x}$MA$_{1-x}$PbI$_{3-x}$Br$_{x}$ perovskites for
  improved carrier collection. \emph{Materials Research Express} \textbf{2019},
  \emph{6}, 085519\relax
\mciteBstWouldAddEndPuncttrue
\mciteSetBstMidEndSepPunct{\mcitedefaultmidpunct}
{\mcitedefaultendpunct}{\mcitedefaultseppunct}\relax
\EndOfBibitem
\bibitem[Premkumar \latin{et~al.}(2019)Premkumar, Kundu, and
  Umapathy]{Premkumar2019impact}
Premkumar,~S.; Kundu,~K.; Umapathy,~S. Impact of cesium in methylammonium lead
  bromide perovskites: insights into the microstructures{,} stability and
  photophysical properties. \emph{Nanoscale} \textbf{2019}, \emph{11},
  10292--10305\relax
\mciteBstWouldAddEndPuncttrue
\mciteSetBstMidEndSepPunct{\mcitedefaultmidpunct}
{\mcitedefaultendpunct}{\mcitedefaultseppunct}\relax
\EndOfBibitem
\bibitem[Jones \latin{et~al.}(2019)Jones, Osherov, Alsari, Sponseller, Duck,
  Jung, Settens, Niroui, Brenes, Stan, Li, Abdi-Jalebi, Tamura, Macdonald,
  Burghammer, Friend, Bulović, Walsh, Wilson, Lilliu, and
  Stranks]{Jones2019lattice}
Jones,~T.~W. \latin{et~al.}  Lattice strain causes non-radiative losses in
  halide perovskites. \emph{Energy Environ. Sci.} \textbf{2019}, \emph{12},
  596--606\relax
\mciteBstWouldAddEndPuncttrue
\mciteSetBstMidEndSepPunct{\mcitedefaultmidpunct}
{\mcitedefaultendpunct}{\mcitedefaultseppunct}\relax
\EndOfBibitem
\bibitem[Grotz \latin{et~al.}(2012)Grotz, Hauf, Dankerl, Naydenov, Pezzagna,
  Meijer, Jelezko, Wrachtrup, Stutzmann, Reinhard, and
  Garrido]{GrotzCharge2012}
Grotz,~B.; Hauf,~M.~V.; Dankerl,~M.; Naydenov,~B.; Pezzagna,~S.; Meijer,~J.;
  Jelezko,~F.; Wrachtrup,~J.; Stutzmann,~M.; Reinhard,~F.; Garrido,~J.~A.
  Charge state manipulation of qubits in diamond. \emph{Nature Communications}
  \textbf{2012}, \emph{3}, 729\relax
\mciteBstWouldAddEndPuncttrue
\mciteSetBstMidEndSepPunct{\mcitedefaultmidpunct}
{\mcitedefaultendpunct}{\mcitedefaultseppunct}\relax
\EndOfBibitem
\bibitem[Walsh and Zunger(2017)Walsh, and Zunger]{Walsh2017instilling}
Walsh,~A.; Zunger,~A. Instilling defect tolerance in new compounds.
  \emph{Nature Materials} \textbf{2017}, \emph{16}, 964--967\relax
\mciteBstWouldAddEndPuncttrue
\mciteSetBstMidEndSepPunct{\mcitedefaultmidpunct}
{\mcitedefaultendpunct}{\mcitedefaultseppunct}\relax
\EndOfBibitem
\bibitem[Jiao \latin{et~al.}(2021)Jiao, Yi, Wang, Li, Hao, Pan, Shi, Li, Liu,
  Zhang, Gao, Zhao, and Lu]{Jiao2021strain}
Jiao,~Y.; Yi,~S.; Wang,~H.; Li,~B.; Hao,~W.; Pan,~L.; Shi,~Y.; Li,~X.; Liu,~P.;
  Zhang,~H.; Gao,~C.; Zhao,~J.; Lu,~J. Strain Engineering of Metal Halide
  Perovskites on Coupling Anisotropic Behaviors. \emph{Advanced Functional
  Materials} \textbf{2021}, \emph{31}, 2006243\relax
\mciteBstWouldAddEndPuncttrue
\mciteSetBstMidEndSepPunct{\mcitedefaultmidpunct}
{\mcitedefaultendpunct}{\mcitedefaultseppunct}\relax
\EndOfBibitem
\bibitem[Prakash \latin{et~al.}(2021)Prakash, Wang, Bucsek, Truttmann, Fali,
  Cotrufo, Yun, Kim, Ryan, Mkhoyan, Alù, Abate, James, and
  Jalan]{Prakash2021self}
Prakash,~A.; Wang,~T.; Bucsek,~A.; Truttmann,~T.~K.; Fali,~A.; Cotrufo,~M.;
  Yun,~H.; Kim,~J.-W.; Ryan,~P.~J.; Mkhoyan,~K.~A.; Alù,~A.; Abate,~Y.;
  James,~R.~D.; Jalan,~B. Self-Assembled Periodic Nanostructures Using
  Martensitic Phase Transformations. \emph{Nano Letters} \textbf{2021},
  \emph{21}, 1246--1252\relax
\mciteBstWouldAddEndPuncttrue
\mciteSetBstMidEndSepPunct{\mcitedefaultmidpunct}
{\mcitedefaultendpunct}{\mcitedefaultseppunct}\relax
\EndOfBibitem
\end{mcitethebibliography}


\begin{thebibliography}{1}

\bibitem{Pieface}
{\sc Cumby, J., and Attfield, J.~P.}
\newblock Ellipsoidal analysis of coordination polyhedra.
\newblock {\em Nature Communications 8}, 1 (Feb 2017), 14235.

\bibitem{PyProcar}
{\sc Herath, U., Tavadze, P., He, X., Bousquet, E., Singh, S., Muñoz, F., and
  Romero, A.~H.}
\newblock Pyprocar: A python library for electronic structure
  pre/post-processing.
\newblock {\em Computer Physics Communications 251\/} (2020), 107080.

\bibitem{Heyd2003hybrid}
{\sc Heyd, J., Scuseria, G.~E., and Ernzerhof, M.}
\newblock Hybrid functionals based on a screened coulomb potential.
\newblock {\em The Journal of Chemical Physics 118}, 18 (2003), 8207--8215.

\bibitem{ASE}
{\sc Larsen, A.~H., Mortensen, J.~J., Blomqvist, J., Castelli, I.~E.,
  Christensen, R., Dułak, M., Friis, J., Groves, M.~N., Hammer, B., Hargus,
  C., Hermes, E.~D., Jennings, P.~C., Jensen, P.~B., Kermode, J., Kitchin,
  J.~R., Kolsbjerg, E.~L., Kubal, J., Kaasbjerg, K., Lysgaard, S., Maronsson,
  J.~B., Maxson, T., Olsen, T., Pastewka, L., Peterson, A., Rostgaard, C.,
  Schiøtz, J., Schütt, O., Strange, M., Thygesen, K.~S., Vegge, T.,
  Vilhelmsen, L., Walter, M., Zeng, Z., and Jacobsen, K.~W.}
\newblock The atomic simulation environment—a python library for working with
  atoms.
\newblock {\em Journal of Physics: Condensed Matter 29}, 27 (2017), 273002.

\bibitem{VESTA}
{\sc Momma, K., and Izumi, F.}
\newblock {{\it VESTA3} for three-dimensional visualization of crystal,
  volumetric and morphology data}.
\newblock {\em Journal of Applied Crystallography 44}, 6 (Dec 2011),
  1272--1276.

\bibitem{stokes2006isodisplace}
{\sc Stokes, H.~T., Hatch, D.~M., Campbell, B.~J., and Tanner, D.~E.}
\newblock {ISODISPLACE}: a web-based tool for exploring structural distortions.
\newblock {\em Journal of Applied Crystallography 39}, 4 (2006), 607--614.

\end{thebibliography}

\end{document}


\date{}
\maketitle

\section{DFT calculated energies}

In Table \ref{DFTenergies} we present the DFT calculated energies used in this work. Kabsch interpolation as outlined in the main text was used to generate the intermediate structures. All values are given for calculations using the screened-exchange HSE06 functional with $\alpha = 0.43$ and spin–orbit coupling \cite{Heyd2003hybrid}. For more information please see the methods section in the main text.

\begin{table}[h]
\caption{DFT calculated energies (in eV) for the materials studied in this work. $E_\mathrm{b}$ is the total energy of the pristine (defect-free) supercell. $E_\mathrm{d}(0)$ and  $E_\mathrm{d}(-)$ are the total energies of the defect supercell in the neutral and negative charge states (not including any charged defect corrections), respectively. $E_\mathrm{d}(-)\rvert_{0}$ is the total energy of the negative charge state evaluated at the equilibrium geometry of the neutral charge state. $E_\mathrm{corr.}$ is the image-charge correction for the negatively charged defect. $\hbar\omega_0$ and $\hbar\omega_-$ describe the potential energy surface (PES) curvature for the neutral and negative charge states, respectively. They assume a harmonic PES, which is only valid near equilibrium geometries.  $\Delta E_\mathrm{p}$ and  $\Delta E_\mathrm{n}$ are the energetic barriers for hole capture at I$_\mathrm{i}^-$ and electron capture at I$_\mathrm{i}^0$ respectively. \label{DFTenergies}}%
\begin{tabularx}{\columnwidth}{l YYYY}
\toprule
 & \ce{MAPbI3} & \ce{MA_{0.875}Cs_{0.125}PbI3} & \ce{MA_{0.75}Cs_{0.25}PbI3}  & \ce{MA_{0.5}Cs_{0.5}PbI3} \\
\midrule
$E_\mathrm{b}$ & -1077.90 & -982.71 & -887.56 & -697.17 \\	
$E_\mathrm{d}(0)$ & -1080.38 &  -984.49 & -889.23 & -698.84 \\ 
$E_\mathrm{d}(-)$ & -1079.55 & -983.42 & -888.00 & -697.68 \\ 
$E_\mathrm{d}(-)\rvert_{0}$ & -1077.57 & -982.33 & -886.89 & -696.45\\ 
$\epsilon_\mathrm{VBM}$ & 1.01 & 1.06  & 1.10 & 1.20 \\ 
$\epsilon_\mathrm{CBM}$ & 2.50 & 2.68 & 2.87 & 2.89 \\ 
$E_\mathrm{corr.}$ & -0.057 & 0.044 & 6E-3 & 0.057 \\
$\hbar\omega_0$ & 5.3E-3 & 4.3E-3 & - & - \\ 
$\hbar\omega_-$ & 7.2E-3 & 6.2E-3 & -  & - \\ 
$\Delta E_\mathrm{p}$ & 0.63 & 0.23 & - & - \\
$\Delta E_\mathrm{n}$ &  0.025 & 0.045 & - & - \\
\end{tabularx}
\end{table}

The charge transition level $\epsilon (q/q\prime)$, reorganisation energy $\lambda$ and Franck-Condon vertical transition energy $E_\mathrm{abs}$ were calculated from the values in the table above, using the expressions given in the main text.

\newpage

\section{Charge localisation at the H-centre defect}
\begin{figure}[h]
\centering
\includegraphics[scale=0.3]{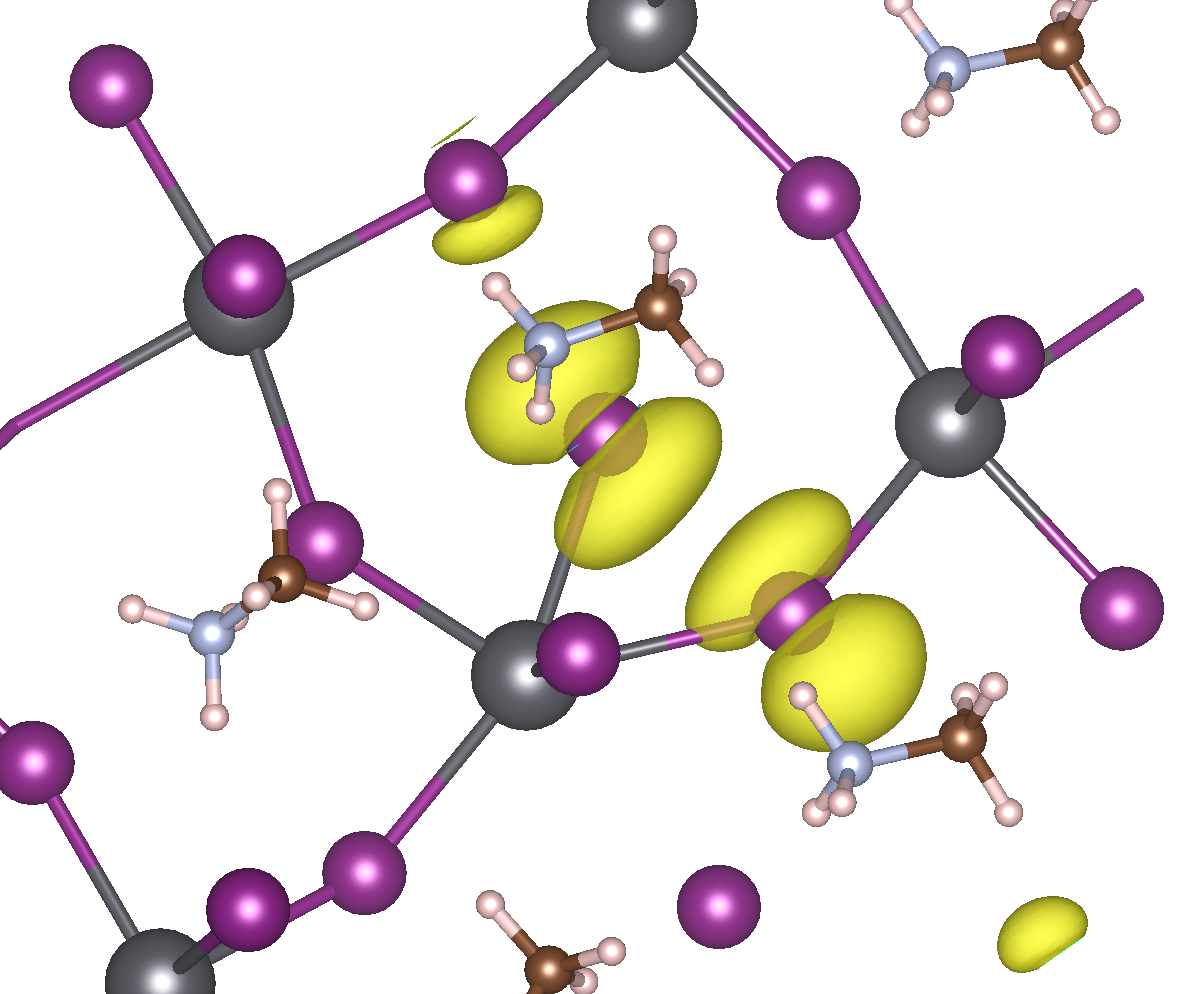}
\caption{Charge density plot for \ce{MA_{0.875}Cs_{0.125}PbI3} with a neutral iodine interstitial defect in the H-centre configuration (I$_2^-$). The plot has been visualised using \texttt{vesta} with an isosurface level of 3.8E-4 \cite{VESTA}. For clarity only part of the full supercell crystal structure is shown. }
\end{figure}

\section{Analysis of the pristine crystal structures}

In Table \ref{MixedCationTable} we present a summary of geometric measures for the \ce{MA_{1-x}Cs_xPbI3} materials in this work. All crystal structures were relaxed using DFT and the pbesol exchange-correlation functional; for more information please see the methods section in the main text. Bond angles were measured using \texttt{ASE} \cite{ASE}. Ellipsoidal properties were calculated using \texttt{Pieface} \cite{Pieface}.

\begin{table}[ht]
\caption{Geometric measures for \ce{MA_{1-x}Cs_xPbI3} materials. The first column specifies the proportion of Cs on the A-site. The second column gives the total volume per formula unit. The third column specifies the mean average volume of the 16 minimum bounding ellipsoid fitted to each inorganic \ce{PbI6} octahedra. The fourth column specifies the mean average of the 48 Pb-I-Pb bond angles. The fifth column specifies the mean average shape factor of the 16 minimum bounding ellipsoids, where a perfect sphere has a shape factor of zero; for more details please see Reference \cite{Pieface}. Standard deviations are given in brackets. In the case of $x=0$ there is only one ellipsoid and three Pb-I-Pb bond angles to consider. \label{MixedCationTable}}%
\begin{tabularx}{\columnwidth}{l YYYYY}
\toprule
x & V$_\mathrm{T}$ ($\angstrom^3$) & $|\mathrm{V}_\mathrm{ellips.}|$ ($\angstrom^3$) & $|\measuredangle_\mathrm{Pb-I-Pb}|$ (deg.)   & $|\Delta\mathrm{S}|$  \\
\midrule
0 & 250.1 & 132.7  &  168.2 ($\pm$2.8) &  0.0047 \\
0.125 & 245.2 & 134.0 ($\pm$0.25) & 161.9 ($\pm$8.1) & 0.0042 ($\pm$0.0081) \\
0.25 & 240.3 & 135.1 ($\pm$0.47)& 156.2 ($\pm$7.7)  & -0.0046 ($\pm$0.0274) \\
0.5 & 239.8 & 134.1 ($\pm$0.66) & 157.6 ($\pm$9.0)&  0.0050 ($\pm$0.0168)\\
\end{tabularx}
\end{table}

\newpage
\section{Analysis of lattice distortions after electron capture at I$^0$}

In Table \ref{ChargeCaptureTable} we present a summary of the lattice distortions induced after charge capture at an iodine interstitial in \ce{MA_{1-x}Cs_xPbI3} materials. All crystal structures were relaxed using DFT and the pbesol exchange-correlation functional; for more information please see the methods section in the main text. Bond angles and bond lengths were measured using \texttt{ASE} \cite{ASE}.

\begin{table*}[h]
\caption{A summary of lattice distortions after charge capture at an iodine interstitial in \ce{MA_{1-x}Cs_xPbI3} materials, calculated using DFT. The first column specifies the proportion of Cs on the A-site.  Columns 2--4 give $\Delta Q =\sqrt{\sum_i m_i \Delta r_{i}^{2}}$ values calculated for all atomic species, the inorganic Pb-I framework and the A-site species. Column 5 gives the mean average absolute change of the 48 Pb-I-Pb bond angles after charge capture. Column 6 gives the mean average absolute change of the 48 Pb-I bond lengths after charge capture. In the case of $x=0$ there are only three Pb-I bond lengths and three Pb-I-Pb bond angles to consider. \label{ChargeCaptureTable}}
\begin{tabularx}{\textwidth}{l YYYYYYYYY}
\toprule
x  & $\Delta\mathrm{Q}_\mathrm{TOTAL}$ & $\Delta\mathrm{Q}_\mathrm{PbI}$ & $\Delta\mathrm{Q}_\mathrm{A}$ & $|\Delta\measuredangle_\mathrm{Pb-I-Pb}|$ & $|\Delta\mathrm{r}_\mathrm{Pb-I}|$   \\
\midrule
0  & 36.77 &  35.16 & 10.75 & 6.03 ($\pm$7.03) &   0.054 ($\pm$0.103) \\
0.125  & 25.94  & 25.09 & 6.612 &  4.78 ($\pm$5.01) &  0.041 ($\pm$0.080) \\
0.25  & 26.30  & 23.17 & 12.46 & 3.61 ($\pm$4.74) &  0.035 ($\pm$0.070)\\
0.5  & 28.96   & 26.62 & 11.42 & 3.77 ($\pm$3.63) &  0.028 ($\pm$0.030)\\
\end{tabularx}
\end{table*}

\section{Electronic density of states}

\begin{figure}[h]
\includegraphics[scale=0.5]{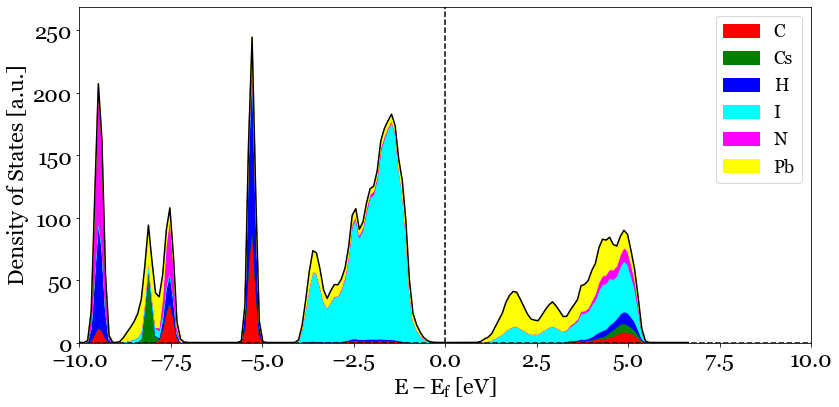}
\caption{Electronic partial density of states analysis for \ce{MA_{0.875}Cs_{0.125}PbI3}, calculated using DFT. Electronic bands from a supercell calculation were unfolded using \texttt{PyProcar} \cite{PyProcar}.}
\end{figure}

\newpage

\section{Symmetry Mode Analysis}

The mode amplitudes are given in $\AA$, and all phonon modes with a displacive mode amplitude above 0.1$\AA$ are shown. The reference (parent) structure is perovskite in the cubic $Pm\bar{3}m$ phase. The rotational motion of the MA molecule was not considered in this analysis; all A-sites were modelled as point particles. This mode amplitudes were analysed using \texttt{Isodistort} \cite{stokes2006isodisplace}.

\begin{figure}[h]
\includegraphics[scale=0.35]{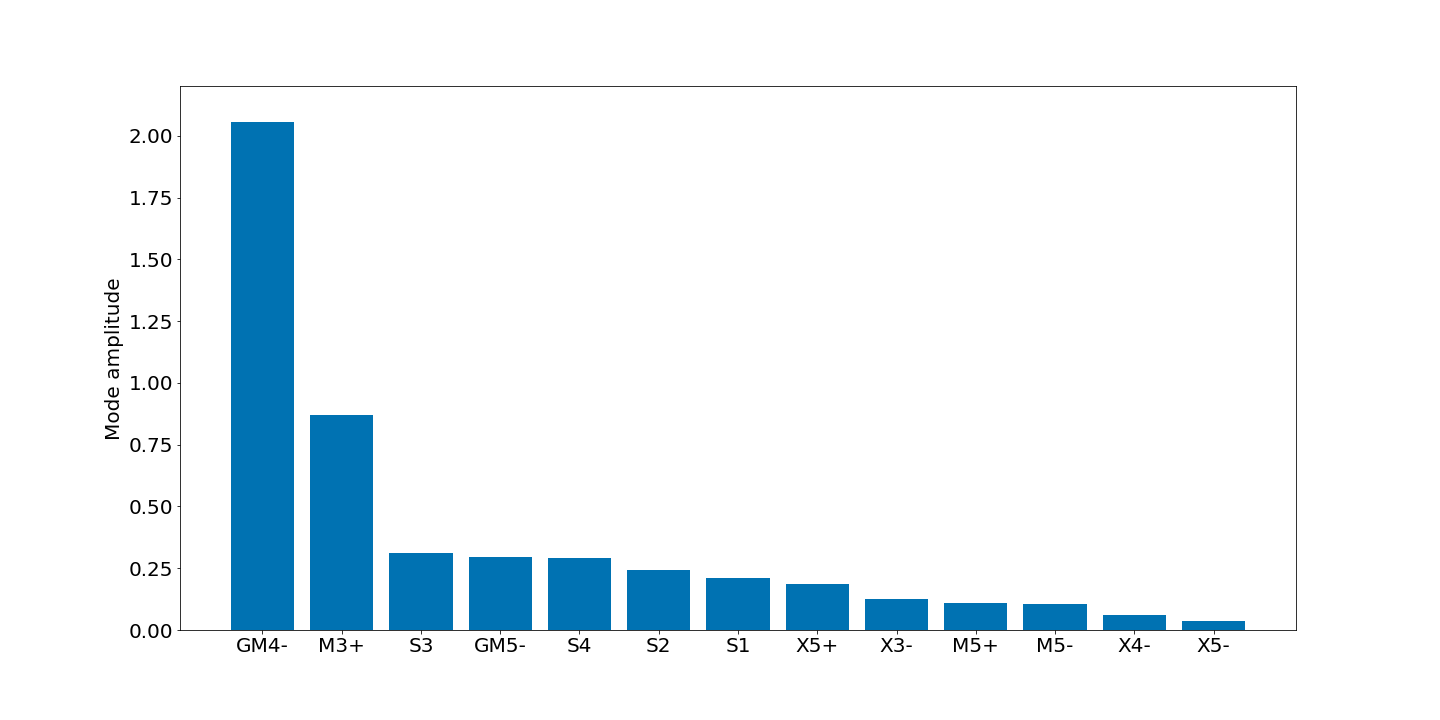}
\caption{Symmetry mode analysis for \ce{MA_{0.875}Cs_{0.125}PbI3}.}
\end{figure}

\begin{figure}[h]
\includegraphics[scale=0.35]{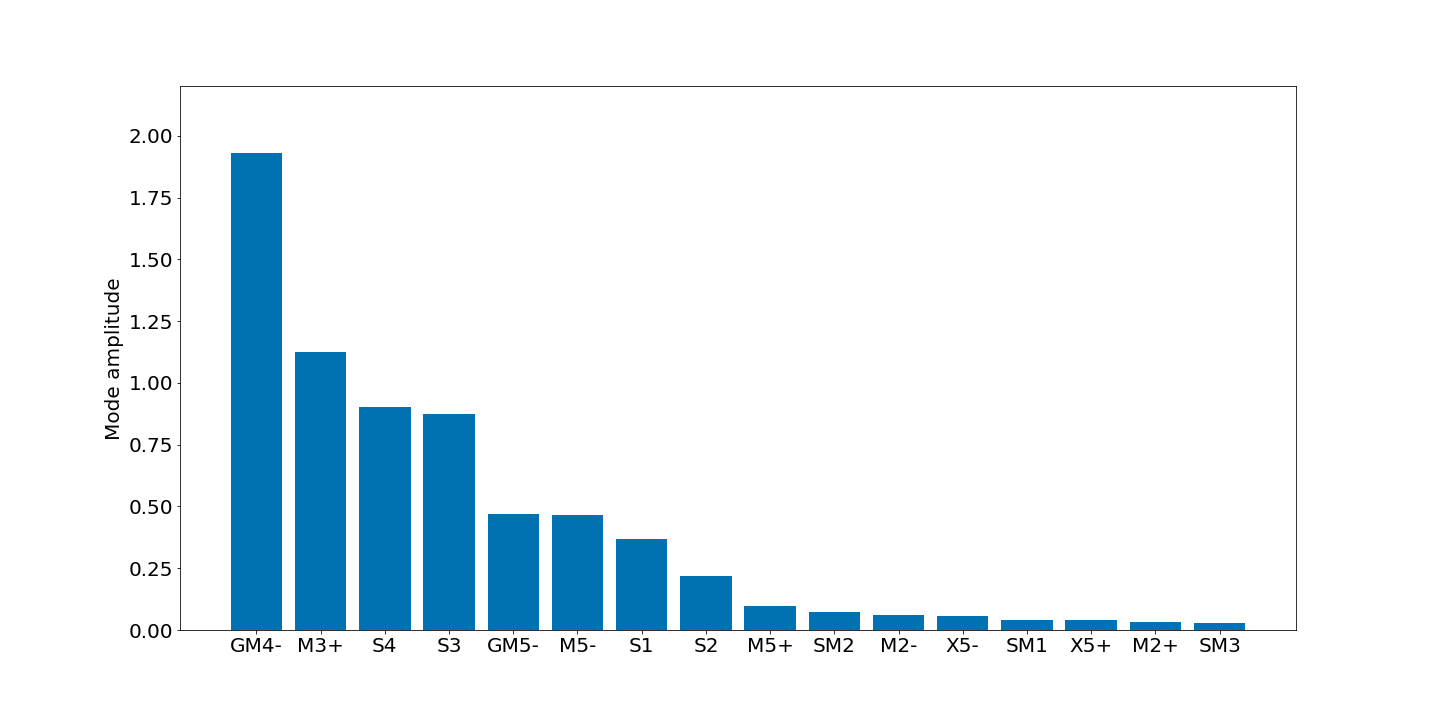}
\caption{Symmetry mode analysis for \ce{MA_{0.75}Cs_{0.25}PbI3}.}
\end{figure}

\begin{figure}[h]
\includegraphics[scale=0.35]{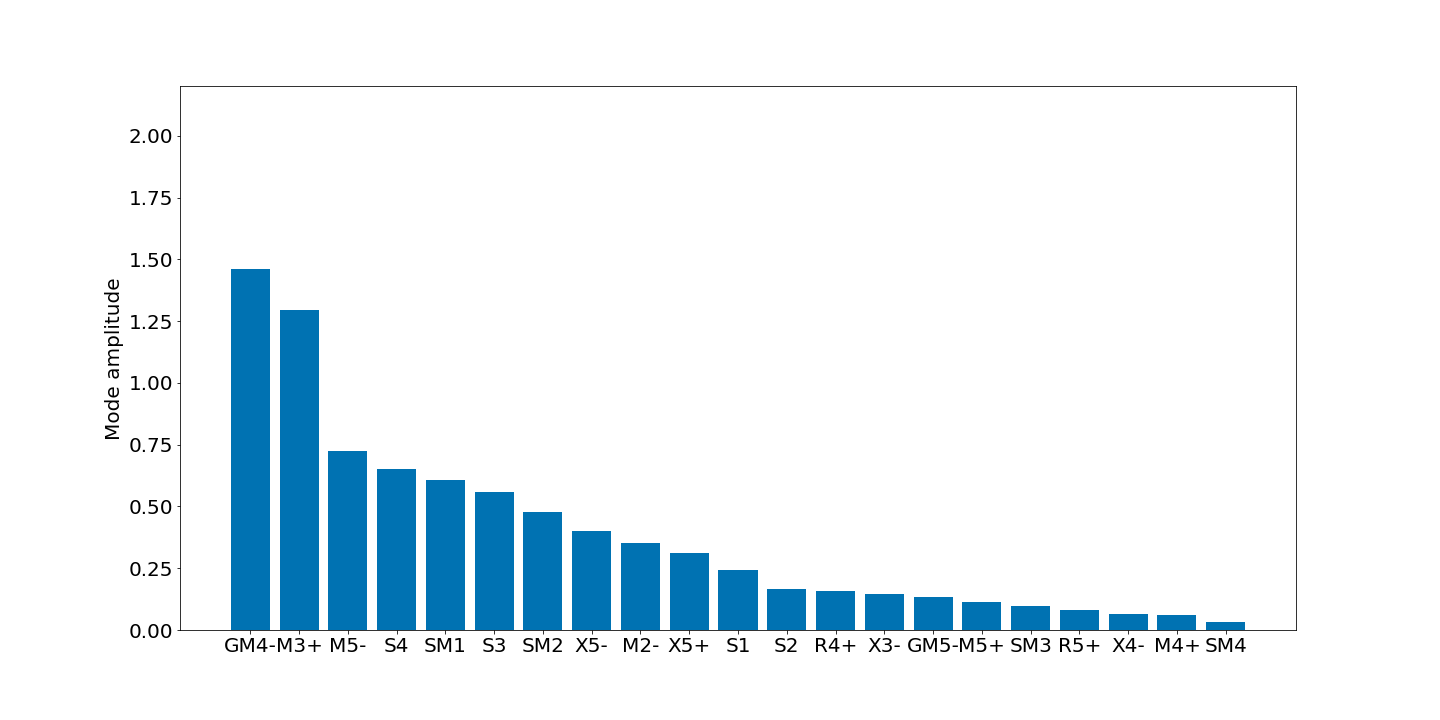}
\caption{Symmetry mode analysis for \ce{MA_{0.5}Cs_{0.5}PbI3}.}
\end{figure}

\bibliography{bibliography}